\documentclass[3p,11pt]{elsarticle}

\usepackage{hyperref}

\journal{Journal}

\usepackage{subcaption} 
\usepackage{amsmath} 
\usepackage{amssymb}
\usepackage{gensymb}
\usepackage{float} 
\usepackage{xcolor}
\usepackage{pgf}
\usepackage{soul}
\usepackage{booktabs}
\usepackage[noabbrev,nameinlink]{cleveref}
\usepackage{bm}

\begin{document}
	

\newcommand{\latticeconstant}{a}                                                
\newcommand{\latticespacing}{d}                                                 
\newcommand{\burgersvector}{b}                                                  
\newcommand{\schmid}{m}                                                         
\newcommand{\rss}{\tau}                                                         
\newcommand{\crss}{s}                                                           
\newcommand{\ratesensitivity}{r}                                                
\newcommand{\hardeningrate}{k_0}                                                
\newcommand{\hardeningratecp}{h_0}
\newcommand{\saturationcrss}{s_\infty}                                          
\newcommand{\hardeningexp}{a}                                                   
\newcommand{\stress}{\sigma}                                                    
\newcommand{\strain}{\varepsilon}												
\newcommand{\youngs}{E}                                                         
\newcommand{\shearmodulus}{G}                                                   
\newcommand{\length}{L}                                                         
\newcommand{\slipangle}{\theta}                                                 
\newcommand{\planeangle}{\phi}                                                  
\newcommand{\displacement}{u}                                                   
\newcommand{\slipstep}{v}                                                       
\newcommand{\slipvelocity}{\dot{\slipstep}}                                     
\newcommand{\refslipvelocity}{\slipvelocity_0}                                  
\newcommand{\sourcelength}{\lambda}                                             
\newcommand{\sourcelengthpdf}{p(\sourcelength)}                                 
\newcommand{\radius}{R}                                                         
\newcommand{\nucleationstress}{s_\text{nuc}}                                    
\newcommand{\minnucleationstress}{\nucleationstress^\text{min}}                 
\newcommand{\maxnucleationstress}{\nucleationstress^\text{max}}                 
\newcommand{\sourcenucleationpdf}{g^+(\nucleationstress)}                         
\newcommand{\nucleationpdf}{p(\nucleationstress)}                               
\newcommand{\nucleationcdf}{P(\nucleationstress)}                               
\newcommand{\minnucleationpdf}{p_\text{min}(\nucleationstress)}                 
\newcommand{\sourcechance}{f_\text{src}}                                        
\newcommand{\edgenucleationpdf}{g^-(\nucleationstress)}                           
\newcommand{\latticefriction}{s_\text{fric}}                                    
\newcommand{\dislocationdensity}{\rho_\text{dis}}                               
\newcommand{\shearstrain}{\gamma}                                               
\newcommand{\shearrate}{\dot{\shearstrain}}
\newcommand{\refshearrate}{\shearrate_0}
\newcommand{\defgradient}{\mathbf{F}}                                           
\newcommand{\velgradient}{\mathbf{L}}                                           
\newcommand{\slipdirection}{\vec{s}_0}                                          
\newcommand{\slipplanenormal}{\vec{n}_0}                                        
\newcommand{\mean}{\mu}                                                         
\newcommand{\stddeviation}{\sigma}                                              
\newcommand{\bandwidth}{l}                                                      
\newcommand{\sourcedensity}{\rho_\text{src}}									

\newcommand{\identity}{\mathbf{I}}
\newcommand{\secondpiolaelastic}{\mathbf{S}_e}                                  
\newcommand{\defgrad}{\mathbf{F}}												
\newcommand{\defgradelastic}{\defgrad_e}										
\newcommand{\defgradplastic}{\defgrad_p}										
\newcommand{\defgradplasticdot}{\dot{\defgrad}_p}								
\newcommand{\velgrad}{\mathbf{L}}												
\newcommand{\velgradelastic}{\velgrad_e}										
\newcommand{\velgradplastic}{\velgrad_p}										
\newcommand{\greenlagrange}{\mathbf{E}}                                         
\newcommand{\elasticitytensor}{\mathbb{C}}                                  	
\newcommand{\schmidtensor}{\mathbf{P}_0}										
\newcommand{\nonschmidtensor}{\mathbf{P}_\text{NS}} 							
\newcommand{\projectionplanenormal}{\vec{n}{'}_0}								
\newcommand{\nonschmidcoeff}{a}												    
\newcommand{\nonschmidstress}{\tau_{\text{NS}}} 								
\newcommand{\rssyield}{\tau_y}													
\newcommand{\slipfamilyA}{\{110\}\left<111\right>}								
\newcommand{\slipfamilyB}{\{112\}\left<111\right>}								
\newcommand{\slipfamilyC}{\{123\}\left<111\right>}								
\newcommand{\planefamilyA}{\{110\}}
\newcommand{\planefamilyB}{\{112\}}
\newcommand{\planefamilyC}{\{123\}}
\newcommand{\compositefactor}{c}												
\newcommand{\mrsspstress}{\tau^*}												
\newcommand{\segmentlength}{L_\text{seg}}										

\newcommand{\potential}{\Pi} 													
\newcommand{\coordvec}{\vec{x}}
\newcommand{\dispvec}{\vec{\displacement}}										
\newcommand{\dispvecmacro}{\dispvec_\text{M}}
\newcommand{\dispvecmean}{\bar{\dispvec}}
\newcommand{\normal}{\vec{n}}                                                   
\newcommand{\traction}{\vec{t}}                                                 
\newcommand{\tractionstiffness}{E^*}											
\newcommand{\tractionlength}{L^*}											
\newcommand{\domain}{\Omega}													
\newcommand{\domaint}{\Omega^\text{2D}}
\newcommand{\domaino}{{\domain_0}}
\newcommand{\domainot}{{\domain_0^\text{2D}}}
\newcommand{\boundary}{\Gamma}		
\newcommand{\boundaryt}{{\boundary^\text{2D}}}							
\newcommand{\boundaryo}{{\boundary_0}}
\newcommand{\boundaryot}{{\boundary_0^\text{2D}}}
\newcommand{\strainenergy}{W}													
\newcommand{\fluctuation}{\vec{w}}												
\newcommand{\lagmultiplier}{\bm{\lambda}}									
\newcommand{\defgradmacro}{\defgrad_\text{M}}									
\newcommand{\firstpiola}{\mathbf{P}}											
\newcommand{\firstpiolamacro}{\firstpiola_\text{M}}
\newcommand{\piolatraction}{\vec{p}}
\newcommand{\locvalue}{\psi}
\newcommand{\locdistance}{\vec{d}_\locvalue}
\newcommand{\energy}{E}
\newcommand{\ausplanes}{\{111\}_{\gamma}}
\newcommand{\marplanes}{\{110\}_{\alpha'}}
\newcommand{\ausplanesalt}{\{557\}_{\gamma}}
\newcommand{\slipdirectionaus}{\vec{s}_{0,\gamma}}                  
\newcommand{\slipplanenormalaus}{\vec{n}_{0,\gamma}}
\newcommand{\slipdirectionmar}{\vec{s}_{0,\alpha'}}                  
\newcommand{\slipplanenormalmar}{\vec{n}_{0,\alpha'}}  

\begin{frontmatter}

\title{High-resolution numerical-experimental comparison of heterogeneous slip activity in quasi-2D ferrite sheets}

\author{J. Wijnen}

\author{T. Vermeij}

\author{J.P.M. Hoefnagels}

\author{M.G.D. Geers}

\author{R.H.J. Peerlings \corref{corauthor}}
\ead{r.h.j.peerlings@tue.nl}

\address{Department of Mechanical Engineering, Eindhoven University of Technology, 5600 MB Eindhoven, The Netherlands}

\cortext[corauthor]{Corresponding author}

\begin{abstract}
	The role of heterogeneity in the plastic flow of thin ferrite specimens is investigated in this study. This is done through a recently introduced quasi-2D experimental-numerical framework that allows for a quantitative comparison of the deformation fields of metal microstructures between experiments and simulations at a high level of detail and complexity. The method exploits samples that are locally ultra-thin ("2D") and hence have a practically uniform microstructure through their thickness. This allows testing more complex loading conditions compared to uniaxial micromechanical experiments while avoiding the complexity of an unknown subsurface microstructure, which limits comparisons between experiments and simulations in traditional integrated approaches at the level of the polycrystalline microstructure. The present approach enables to study the effect of microstructural features such as grain boundaries. To study the role of stochastic fluctuations, a constitutive model is employed which introduces random heterogeneity into a crystal plasticity model. A detailed analysis of the simulations is performed at the level of individual slip systems. Since both experimental and numerical results are susceptible to stochastic fluctuations, the outcomes of many simulations are compared to the experimentally obtained result. This comparison allows us to determine how a single experiment relates to an ensemble of simulations. Additionally, results obtained with a conventional crystal plasticity model are considered. The analysis reveals that the heterogeneity in the plasticity model is essential for accurately capturing the deformation mechanisms.
\end{abstract}

\begin{keyword}
Crystal plasticity \sep Numerical-experimental \sep Strain localization \sep Plastic heterogeneity \sep Dislocation sources
\end{keyword}

\end{frontmatter}



\section{Introduction}

Plastic deformation in metals is a complex phenomenon, governed by various length scales. Contrary to the apparent homogeneous deformation observed at the macroscale, metal plasticity is stochastic and heterogeneous by nature, introducing fluctuations at various scales. At the smallest scale, plastic deformation of a crystal results from the gliding of dislocations over discrete glide planes. Multiple dislocations move collectively in an avalanche-like motion, resulting in intermittent plastic deformation in the form of strain bursts \cite{Miguel2001,Kapetanou2015,Papanikolaou2018,Weiss2020}. At the scale of single grains, heterogeneities in the dislocation substructure, such as the locations and strengths of dislocation sources, introduce fluctuations and strain localizations in the deformation of metals. Additionally, a heterogeneous microstructure, e.g.\ consisting of different phases, promotes strain localizations at the microstructural scale \cite{Tasan2014,Alaie2015}. The inhomogeneity of plastic deformation at a certain length scale strongly affects strain localizations at larger scales, all the way up to macroscopic localization and failure \cite{Alaie2015,Zheng2020,Silva2023}. Therefore, studying fluctuations at different length scales is essential for improving our understanding of metal plasticity. 

This paper presents a numerical-experimental analysis of the plastic heterogeneity at the grain scale. The main objective is to demonstrate the role of plastic strain localizations in the deformation behavior of grains. Key components that facilitate this study are the adopted 'quasi-2D' experimental methodology, recently developed to enable detailed comparisons between experiments and simulations at this scale \cite{Vermeij2023b}, and the employment of a computational model that introduces stochastic sub-grain fluctuations of plastic properties that naturally promote strain localizations \cite{Wijnen2021}. Additional challenges that obstruct comprehensive experimental-numerical comparisons at this scale are addressed here. The most prominent challenge is how to compare simulations and experiments. Since the precise configuration of the underlying dislocation structure, e.g.\ the locations of dislocation sources, is not known in experiments, obtaining a perfect one-to-one match between simulated and experimentally obtained strain fields is not possible. Instead, representative characteristics of a strain field are computed and are used to compare against an ensemble of strain fields.

In the literature, micro- and nanoscale mechanical testing, commonly performed through the loading of miniaturized tension or compression specimens, has been used extensively to investigate heterogeneous microscale plasticity \cite{Uchic2009,Greer2011,Dehm2018}. These experiments rely on a detailed three-dimensional (3D) characterization of the deformation mechanisms, and enable detailed comparisons with simulations, for example, at the level of individual slip systems \cite{Du2018b,Raabe2007}. Micro- and nanoscale experiments are often accompanied by discrete dislocation (DD) simulations to gain insights into the fundamental deformation mechanisms of the dislocation network, e.g. into the stochastics of strain bursts \cite{Csikor2012,Alcala2020} and dislocation sources \cite{Shishvan2010,Elawady2009,Ryu2015}.

Most microscale experiments are limited in complexity through the microstructure morphology, specimen shape, and loading path due to the small specimen size in at least two directions and the absence of bulk material in these directions. This impedes the investigation of the interaction of deformation mechanisms and the influence of microstructure morphology. To advance our understanding in the direction of the bulk (or sheet) material, integrated experimental-numerical approaches with the same level of detail but with a larger degree of microstructural complexity are desired. On the other hand, in traditional polycrystalline bulk samples, only the surface microstructure can be characterized. Here, the unknown subsurface structure limits the interpretation of the observed deformation and the direct comparison to simulations, since severe two-dimensional (2D) simplifications through the thickness are usually made. 

The recently developed integrated experimental-numerical quasi-2D framework, exploited in this work, aims to bridge the gap between small-scale approaches and traditional integrated approaches at the level of a polycrystal \cite{Vermeij2023b}. The considered specimens have a thickness of at most a few micrometers in the gauge region, while their in-plane dimensions span multiple grains. This results in pancake-shaped grains, which can be characterized on both the front and rear sides of the specimen. Interpolating these surface microstructures through the (small) thickness is a reasonable assumption that results in an accurate 3D representation of the actual microstructure. In this way, direct comparisons between simulations and experiments are possible on specimens that contain several microstructural features, which do not suffer from unknown subsurface effects.

At the considered microstructural scale, strain localizations triggered by heterogeneities in the underlying dislocation network are prominent. To analyze the effect of the heterogeneous strain fields at this scale, computational material models need to be employed that take this heterogeneity into account. However, most numerical modeling frameworks addressing these inhomogeneities, e.g.\ DD simulations, are computationally too expensive for modeling experiments at real scale in full detail. The discrete slip plane (DSP) model, developed in \cite{Wijnen2021}, overcomes this issue by introducing fluctuations of the slip resistance into a crystal plasticity model by considering one of the main characteristics of the underlying dislocation network, i.e.\ the strength and spatial distribution of dislocation sources. A notable advantage of this model is that it can be integrated into a conventional crystal plasticity (CP) model, only requiring a pre-processing step to determine the local material parameters based on the dislocation source stochastics, without directly simulating dislocation (source) interactions. It was shown in an earlier study that including this stochastic behavior can explain the diversity in the many active slip systems observed in uniaxial tensile tests of single-crystal ferrite \cite{Wijnen2023}, contrary to a conventional CP model that does not include any heterogeneity at the scale of single crystals. In this paper, we study the role of stochastic fluctuations in the deformation behavior of quasi-2D specimens. These thin samples contain more microstructural complexity than uniaxial tensile tests since only one of the dimensions is small. The combination of the integrated quasi-2D framework and the DSP model allows to study larger areas that contain microstructural features such as grain boundaries.

As stated before, the stochastic variations in both the experiments and simulations, make it impossible to achieve a perfect match between the two. Additionally, only a single unique quasi-2D sample can be fabricated, i.e.\ an experiment cannot be repeated on the same sample. Therefore, instead of directly comparing deformation fields between experiments and simulations, we aim to focus on characteristics of the deformation field that are statistically representative. For this reason, experimental strain fields are decomposed into contributions from individual slip systems by employing the SSLIP method \cite{Vermeij2023}. Additionally, the degree of heterogeneity in these slip fields is determined. By using these characteristic quantities of the strain fields, we aim to determine how a single experimental sample relates to the ensemble of simulations.

A detailed analysis of two ferrite regions is presented. To investigate the role of plastic heterogeneity on the deformation of grains, differences between experimental observations, simulations with the DSP models, and simulations with a standard (smeared) CP model are studied. The latter accounts for inhomogeneities between different phases and grains through crystallographic texture but lacks sub-grain fluctuations.

This paper has the following structure. In \Cref{section:models}, the computational material models employed are briefly summarized. \Cref{section:methodology} presents two novel aspects extending the quasi-2D integrated experimental-numerical framework. The set of boundary conditions applied to the regions of interest is one of these aspects. The experimentally measured deformations at the edge of the examined region are applied in a weak sense, allowing for deviations from the measured displacements in the simulations. Furthermore, we introduce some characteristic measures of the deformation field, based on which a detailed statistical comparison at the level of individual slip systems is performed. The above-described sections use a simple single-grain ferrite region for demonstration purposes. A region containing a grain boundary is presented as a case study in \Cref{section:case1}. Finally, the conclusions are summarized in \Cref{section:conclusions}


\section{Material models}
\label{section:models}

In this section, the two material models used in simulations throughout this paper are summarized. We first briefly review a conventional phenomenological crystal plasticity (CP) model in \Cref{section:models_cp}, because it serves as a basis for the more advanced discrete slip plane (DSP) model. Furthermore, the standard CP results are used as a classical reference for comparison purposes. In \Cref{section:models_dsp}, we discuss the DSP model, which considers the spatial and strength distribution of dislocation sources. This results in a heterogeneous slip resistance within a grain and consequently in heterogeneous slip distributions. Finally, the adopted model parameters are discussed in \Cref{section:models_parameters}.

\subsection{Crystal plasticity model}
\label{section:models_cp}

Crystal plasticity (CP) models are commonly used to study polycrystalline materials and microstructures in relation to their macroscopic mechanical properties \cite{roters2019}. In the conventional phenomenological CP model employed here, the deformation of a material point is the accumulated (homogeneous) behavior of slip on many underlying atomic slip planes.

The model is formulated in a finite deformation setting, in which the deformation gradient tensor, $\defgrad$, is split into an elastic part, $\defgradelastic$, and a plastic part, $\defgradplastic$, via a multiplicative decomposition:
\begin{equation}
	\defgrad = \defgradelastic \cdot \defgradplastic
	\ . \
\end{equation}
The crystallographic orientation of a grain is taken into account in the plastic velocity gradient by considering the individual contributions of all the slip systems:
\begin{equation}
	\velgradplastic = \sum_{\alpha=1}^{N} \shearrate^{\alpha} \slipdirection^{\:\alpha} \otimes \slipplanenormal^{\:\alpha}
	\ , \
\end{equation}
where $\shearrate^\alpha$ is the shear rate on slip system $\alpha$, which has a slip plane normal $\slipplanenormal^{\:\alpha}$ and slip direction $\slipdirection^{\:\alpha}$. $N$ denotes the total number of slip systems. The shear rate of a slip system is related to the resolved shear stress on that slip system, $\rss^\alpha$, by a phenomenological power law:
\begin{equation}
	\shearrate^\alpha = \refshearrate \left( \frac{|\rss^\alpha|}{\crss^\alpha} \right)^\frac{1}{\ratesensitivity} \text{sign} \left( \rss^\alpha \right)
	\ ,
	\label{eq:shearrate}
\end{equation}
with slip resistance $\crss^\alpha$, reference shear rate $\refshearrate$ and rate-sensitivity exponent $\ratesensitivity$. The slip resistance $\crss^\alpha$ has an initial value $\crss_0$ and its evolution is given by
\begin{equation}
	\dot{\crss}^\alpha = \hardeningratecp \left( 1 - \frac{\crss^\alpha}{\saturationcrss} \right)^a \sum_{\beta=1}^N q^{\alpha \beta} | \shearrate^\beta |
	\ ,
\end{equation}
where $h_0$ and $\alpha$ are hardening parameters and the constants $q^{\alpha \beta}$ characterize the mutual interaction between slip systems, by latent hardening. $q^{\alpha \beta}$ takes a value of 1 when slip systems $\alpha$ and $\beta$ share the same slip plane, and a value of $q_n$ otherwise.

%

\subsection{Discrete slip plane plasticity}
\label{section:models_dsp}

At small scales, the discrete and stochastic nature of plastic slip is more prominent. This is particularly true inside individual grains, where a highly heterogeneous plastic slip activity is generally observed. Conventional CP models capture fields that are much smoother than observed experimentally. For this reason, the so-called discrete slip plane (DSP) model has been proposed in earlier work, see Wijnen et al.\ \cite{Wijnen2021}. Based on the assumption that the heterogeneous plastic deformation is predominantly due to the specific configuration of dislocation sources and obstacles to dislocation glide, the model introduces a spatial variation of slip system properties probed randomly from a stochastic distribution of dislocation sources, which naturally induces more localized slip patterns compared to CP.

In the DSP model, in its fundamental form, all discrete atomic glide planes of a particular slip system in the finite size crystal are considered and the amount of slip on them is characterized by a relative displacement (or disregistry) field $\slipstep$. The initial resistance against slip, i.e\ against the evolution of $\slipstep$ on an atomic plane is given by:
\begin{equation}
	\crss_0^{\alpha,i} = \nucleationstress^{\alpha,i}  + \latticefriction + \tfrac{1}{2} \shearmodulus \burgersvector \sqrt{\dislocationdensity}
	\ ,
	\label{eq:initialcrss}
\end{equation}
where $\nucleationstress$ denotes the nucleation stress, $\latticefriction$ the lattice friction, $\shearmodulus$ the shear modulus, $\burgersvector$ the length of the Burgers vector and $\dislocationdensity$ the initial dislocation density. It is assumed that the nucleation stress varies between individual atomic planes due to the presence of one or multiple dislocation sources (giving a low $\nucleationstress$) or obstacles (giving a high $\nucleationstress$). For each plane, the nucleation stress is randomly sampled from a probability density function (PDF) based on the physics of dislocation sources. Most planes are assigned a high nucleation stress, on the order of the theoretical shear strength of the crystal, representing planes that do not contain a dislocation source. The nucleation stress of planes that do contain a dislocation source is based on the strength of a single-arm (SA) source.

\begin{figure}[!tb]
	\centering
	\begin{subfigure}{\linewidth}
		\includegraphics[width=\linewidth]{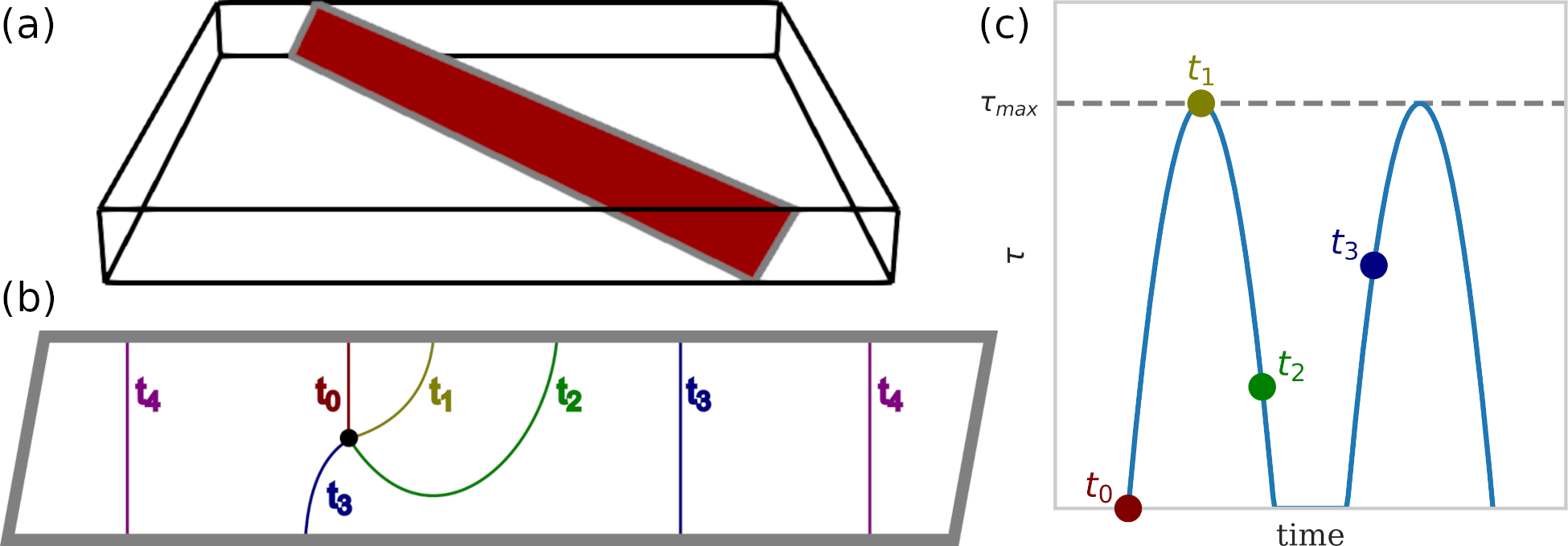}
		\phantomcaption \label{fig:sa_source_plate_a}
		\phantomcaption \label{fig:sa_source_plate_b}
		\phantomcaption \label{fig:sa_source_plate_c}
	\end{subfigure}
	\caption{Schematic of the cross section of a thin plate containing a SA source. (a) Shows the outline of the thin plate with the cross section plane depicted in red. (b) Shows a perpendicular view of such a plane containing a SA source. The mechanism of a SA source in a thin plate is depicted by configurations $t_1$ through $t_4$. (c) A sketch of the shear stress required during the operation of the SA source.}
	\label{fig7:sa_source_plate}
\end{figure}

SA sources are considered to be the dominant plastic mechanism in micropillar compression tests or microtensile tests \cite{Uchic2009,Kiener2011,Samaee2018}. In such geometries, a SA source spirals around its pinning point, resulting in a plastic slip on its plane. The mechanism of a SA source in a thin plate or film slightly deviates from a SA source in a micropillar. \Cref{fig:sa_source_plate_a} shows a schematic of a thin plate with a single plane of a slip system displayed in red. \Cref{fig:sa_source_plate_b} shows the perpendicular view of the plane, containing a SA dislocation source. Additionally, \Cref{fig:sa_source_plate_c} shows a schematic of the resolved shear stress required during the operation of the source. Initially, at configuration $t_0$, the dislocation in \Cref{fig:sa_source_plate_b} is pinned to a point in the center of the plate. The other end of the dislocation is connected to the free surface at the top of the plate. When a resolved shear stress is applied to the glide plane of the dislocation, it bows out. The stress that bows out the dislocation increases until it reaches its critical configuration, $t_1$, corresponding to a resolved shear stress denoted as $\tau_\text{max}$ in \Cref{fig:sa_source_plate_c}. In this configuration, the dislocation has the maximum curvature, and, consequently, line tension that needs to be overcome. If the resolved shear stress is sufficiently high to overcome this critical configuration, it continues to bow out (configuration $t_2$). Finally, part of the dislocation reaches the free surface at the bottom of the plate and the dislocation is split in two, as shown in configuration $t_3$. One dislocation segment is still connected to the pinning point while its other end is connected to the free surface at the bottom of the plate. The other new dislocation is connected to the free surfaces on both sides of the plate. This process repeats itself on the other side of the pinning point, where it creates another dislocation, shown in configuration $t_4$. The pinned dislocation is now back in its initial configuration, $t_0$. Therefore, in contrast to a SA source in a micropillar, a SA source in a thin plate generates new dislocations that are connected to the free surfaces at both sides of the plate, which means that they cannot be annihilated at the free surface (top or bottom). 

The required resolved shear stress to activate a SA source is inversely proportional to the length over which the dislocation has to bow out in its critical configuration ($t_1$), i.e.\ the shortest distance on the glide plane from the pinning point to one of the free surfaces. Note that in \Cref{fig:sa_source_plate_c}, the two peaks have an equal height since the pinning point in \Cref{fig:sa_source_plate_b} is located exactly in the middle of the plate. If the pinning point were located closer to the top surface, the height of the first peak would increase, whereas the second peak would become lower. The opposite effect would be observed if the pinning point was closer to the bottom surface.

Therefore, the adopted PDF for $\nucleationstress$ is based on parallelogram-shaped planes for cuboid geometries, as presented in Reference \cite{Wijnen2023}. One dimension of the cuboid geometry is taken equal to the average thickness of a grain, while the other two dimensions are adopted as the in-plane grain size. Note that since the thickness of most grains in the regions of interest is much smaller than their in-plane dimensions, the strength of the SA sources, related to the shortest distance of the pinning point to the free surface, is mainly determined by the thickness.

The kinetics of slip over a single atomic plane, $i$, is described by
\begin{equation}
	\slipvelocity^{\alpha,i} = \refslipvelocity \left( \frac{|\rss^{\alpha,i}|}{\crss^{\alpha,i}} \right) ^\frac{1}{\ratesensitivity} \text{sign}(\rss^{\alpha,i})
	\ ,
	\label{eq:velocity}
\end{equation}
where $\slipvelocity^{\alpha,i}$ and $\refslipvelocity$ are velocities instead of the shear rates appearing in the conventional CP counterpart of this equation, \Cref{eq:shearrate}. The slip resistance of each plane has an initial value $\crss^{\alpha,i}_0$ and evolves asymptotically towards $\crss^{\alpha,i}_\infty$ according to the relationship
\begin{equation}
	\dot{\crss}^{\alpha,i} = \hardeningrate \left(1 - \frac{\crss^{\alpha,i}}{\saturationcrss^{\alpha,i}} \right)^\hardeningexp|\slipvelocity^{\alpha,i}|
	\ ,
	\label{eq:hardening_discrete}
\end{equation}
where $\hardeningrate$ is the initial hardening rate.

Modeling all atomic planes of all the slip systems in a three-dimensional volume of the sizes considered here is computationally untractable. Therefore, atomic slip planes are clustered in bands, in which it is assumed that only the weakest slip plane experiences slip. Dividing this plastic relative displacement over the width of the band, $\bandwidth$, results in a mean shear strain rate within the band given by
\begin{equation}
	\dot{\gamma}^\alpha = \frac{\dot{v}_0}{l} \left( \frac{|\tau^\alpha|}{s^{\alpha,i_\text{min}}} \right)^\frac{1}{r} \text{sign} \left( \tau^\alpha \right)
	\ .
\end{equation}
We employ this relationship to implement the DSP model as a conventional CP model with band-specific material properties, randomly sampled from a stochastic distribution. Furthermore, the initial slip resistance $s_0$ adopted in a band is equal to that of the weakest plane, $s^{\alpha,i_\text{min}}$, i.e.\ the plane with the lowest slip resistance. This approach is taken for all ($N$) considered slip systems of the crystal, with a separate band for each slip system, implying that a particular material point (integration point in a finite element setting) belongs to $N$ bands, each with its own stochastic properties.

As said, the slip resistance in a band is given by the slip resistance of the weakest atomic plane that experiences slip. Therefore, \Cref{eq:hardening_discrete} of the weakest atomic slip plane in a band is rewritten in terms of the mean shear rate in that band, resulting in
\begin{equation}
	\dot{\crss}^{\alpha,i_\text{min}} = \hardeningrate \bandwidth \left( 1 - \frac{\crss^{\alpha,i_\text{min}}}{\saturationcrss^{\alpha,i_\text{min}}} \right)^\hardeningexp \sum_{\beta=1}^N q^{\alpha\beta} |\dot{\shearstrain}^\beta|
	\ .
	\label{eq:hardening}
\end{equation}
Note that latent hardening is additionally introduced into the above equation through constants $q^{\alpha \beta}$.

A more detailed treatment of the DSP model is presented in Wijnen et al.\ \cite{Wijnen2021}.

\subsection{Model parameters}
\label{section:models_parameters}

The parameters for both the conventional CP model and the DSP model have been identified from the same uniaxial microtensile tests on monocrystal ferrite specimens of interstitial free steel, in Du et al.\ \cite{Du2018} and Wijnen et al.\ \cite{Wijnen2023}, respectively. The resulting model parameters, shown in \Cref{tab:ferrite_parameters}, are adopted in the present study.

\begin{table}[!tb]
	\centering
	\caption{Model parameters used for interstitial free ferrite in the CP and DSP model.}
	\label{tab:ferrite_parameters}
	\begin{tabular}{@{}ll|ll@{}}
		\multicolumn{2}{c|}{\textbf{CP}} & \multicolumn{2}{|c}{\textbf{DSP}} \\ \hline
		$\refshearrate$    & 0.001 s$^{-1}$ & $\refslipvelocity$    & 0.1  $\mu$m/s                         \\
		$\ratesensitivity$ & 0.05  & $\ratesensitivity$    & 0.05                          \\
		$\crss_0$          & 15 MPa   & $\bandwidth$          & 1 $\mu$m                            \\
		$\saturationcrss$  & 250 MPa  & $\latticefriction$    & 20 MPa                           \\
		$\hardeningratecp$ & 120 GPa  & $\dislocationdensity$ & 7$\cdot 10^{11}$ m$^{-2}$ \\
		$a$                & 17.5  & $\saturationcrss$     & 2$s_0$                             \\
		$q_n$              & 1.4   & $\hardeningrate$      & 30 GPa/$\mu$m                          \\
		&       & $a$                   &  5                             \\
		&       & $q_n$                 &  1.4                          \\
	\end{tabular}
\end{table}


\section{Novel aspects of the integrated methodology}
\label{section:methodology}

The DSP model introduced in \Cref{section:models_dsp} enables a detailed analysis of the kinematics of metallic microstructures. Unlike conventional CP, the stochastic nature of this model entails heterogeneous and partially localized strain distributions with similar mean characteristics, but generally a different spatial distribution compared to the measured pattern. To deal with this stochastic variability, two additional elements of the quasi-2D integrated experimental-numerical approach of Vermeij et al.\ \cite{Vermeij2023b} are introduced. First, for the applied boundary conditions the displacements extracted from the in-situ SEM-DIC measurements are adopted. However, unlike the earlier method by Vermeij et al.\ \cite{Vermeij2023b}, fluctuations with respect to the measured displacements are allowed, which relax artificial stress concentrations near the boundary of the region of interest (ROI) and allows for strain localization at other locations than in the experiment. Second, two quantitative indicators are introduced to determine how well the experiment matches the ensemble of random realizations of the DSP model.

\begin{figure}[!htb]
	\centering
	\begin{subfigure}{\linewidth}
		\centering
		\includegraphics[width=\linewidth]{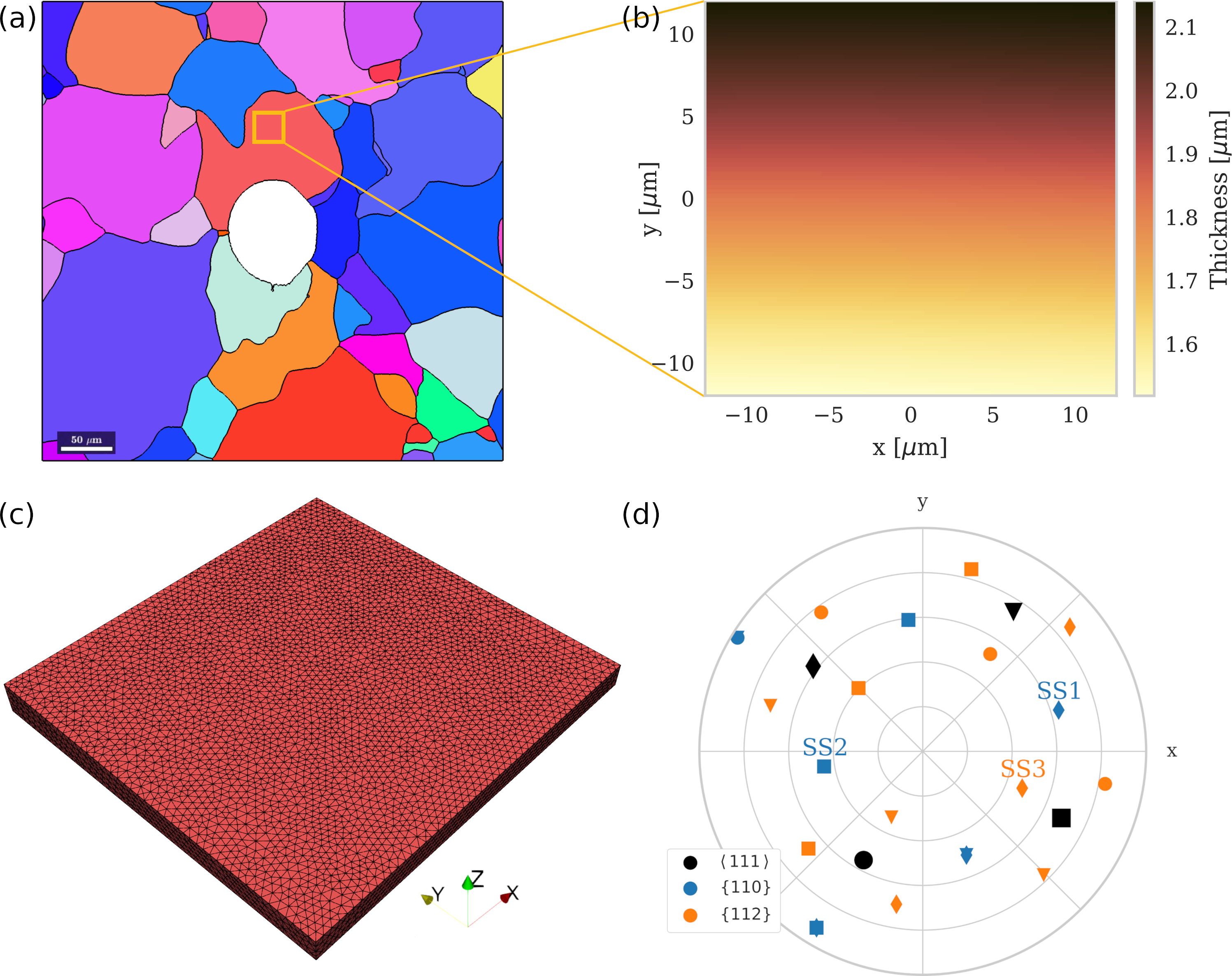}
		\phantomcaption \label{fig:case0_overview_a}
		\phantomcaption \label{fig:case0_overview_b}
		\phantomcaption \label{fig:case0_overview_c}
		\phantomcaption \label{fig:case0_overview_d}
	\end{subfigure}
	\caption{(a) EBSD map of the area around the hole of the specimen. The small ROI is marked with a yellow rectangle. (b) Thickness profile of the ROI. (c) Discretization of the ROI consisting of 124,126 quadratic tetrahedral elements and 186,649 nodes. (d) Pole figure showing the orientations of the slip plane normals and slip directions of the considered grain. The blue and orange markers denote the ${110}$ and ${112}$ plane normal, respectively, while the slip directions are denoted by black markers. The plane normal of a slip system is depicted with the same marker as its slip direction.}
	\label{fig:case0_overview}
\end{figure}

A simple microstructural ROI, i.e.\ a single crystal domain without grain boundaries, is used for demonstration purposes throughout this section. The ROI is part of a larger specimen of which the thickness in the center is reduced by electrolyte polishing until a small hole is created, as described in detail in Reference \cite{Vermeij2023b}. \Cref{fig:case0_overview_a} shows the electron backscatter diffraction (EBSD) image of an area around the hole. The ROI is located in the grain immediately above the hole, as indicated by the yellow rectangle. The thickness profile of the ROI is shown in \Cref{fig:case0_overview_b}, and ranges between 1.5 and 2.1 $\mu$m. The discretization that is used in the simulations, containing 124,126 quadratic tetrahedral elements and 186,649 nodes, and the slip system orientations of the grain are shown in \Cref{fig:case0_overview_c,fig:case0_overview_d}, respectively.

\subsection{Boundary conditions}

\begin{figure}[!ht]
	\centering 
	\begin{subfigure}[b]{.8\linewidth}
		\centering
		\includegraphics[width=\linewidth]{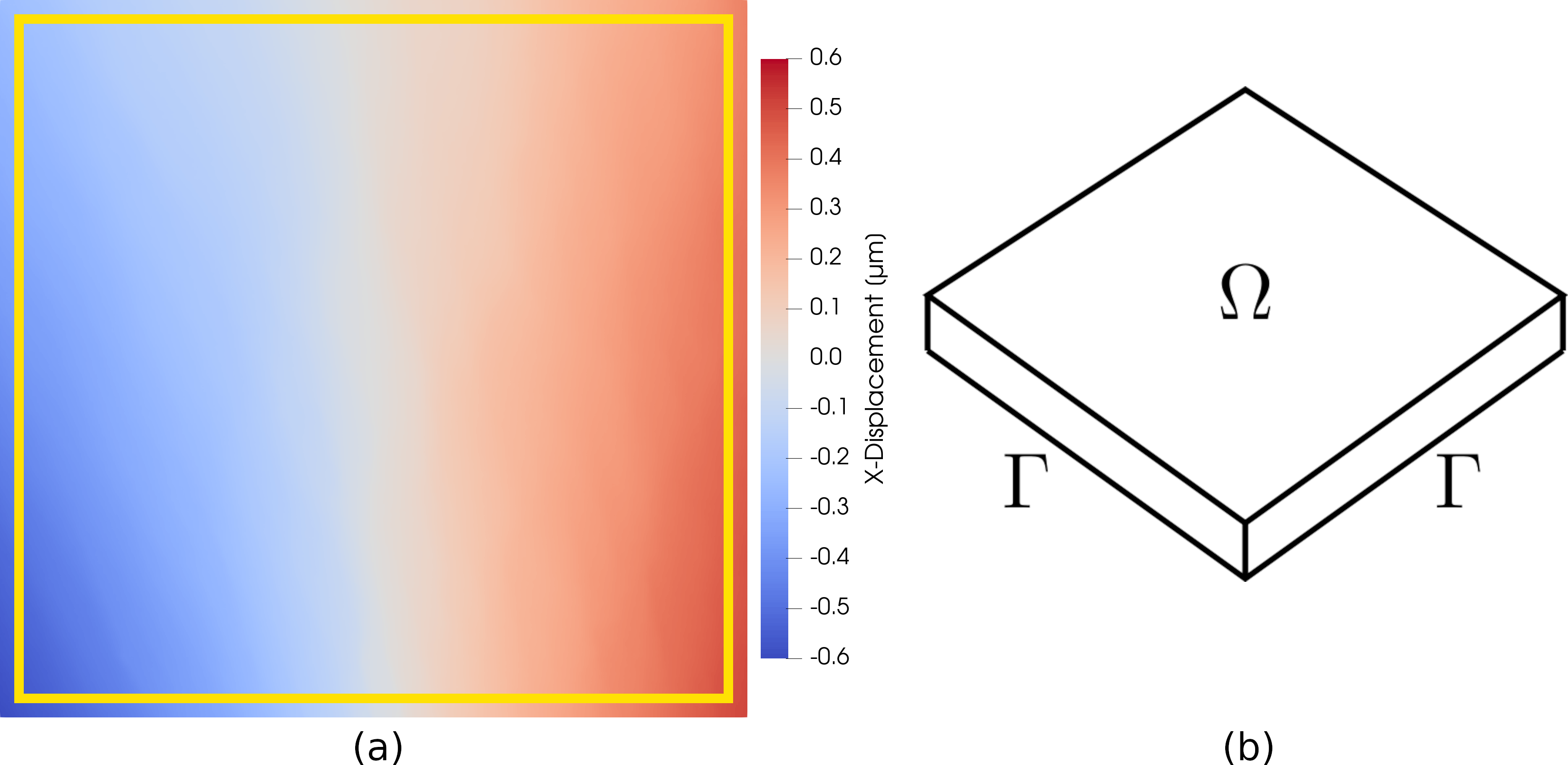}
		\phantomcaption	\label{fig:apply_bcs_a}
		\phantomcaption	\label{fig:apply_bcs_b}
	\end{subfigure}
	\caption{(a) Displacements in the $x$-direction measured with SEM-DIC. The yellow border denotes the region of interest used in the simulations. (b): Schematic of the region of interest. $\domain$ denotes the volume. $\boundary$ denotes the through-thickness boundaries to which the boundary conditions are applied. In reality, these boundaries are connected to surrounding material.}
	\label{fig:apply_bcs}
\end{figure}

A key characteristic of the employed integrated experimental-numerical framework is that the displacements measured on the boundary of the region of interest in the experiment are applied as boundary conditions in the simulations. \Cref{fig:apply_bcs_a} shows the displacements in the x-direction obtained by SEM-DIC from the experiment. The boundary of the region considered in the simulations is depicted with a yellow line. In the experiment, only the surface x- and y-displacements are measured in the region. However, the simulations are conducted in 3D. Therefore, the measured in-plane displacements at the ROI edges are extruded through the thickness, i.e.\ they are prescribed to the full boundary denoted by $\Gamma$ in the sketch of \Cref{fig:apply_bcs_b}. This boundary condition set is referred to as fully prescribed boundary conditions (FPBCs) in the remainder of this section.

\begin{figure}[!tb]
	\begin{subfigure}{\linewidth}
		\centering
		\includegraphics[width=\linewidth]{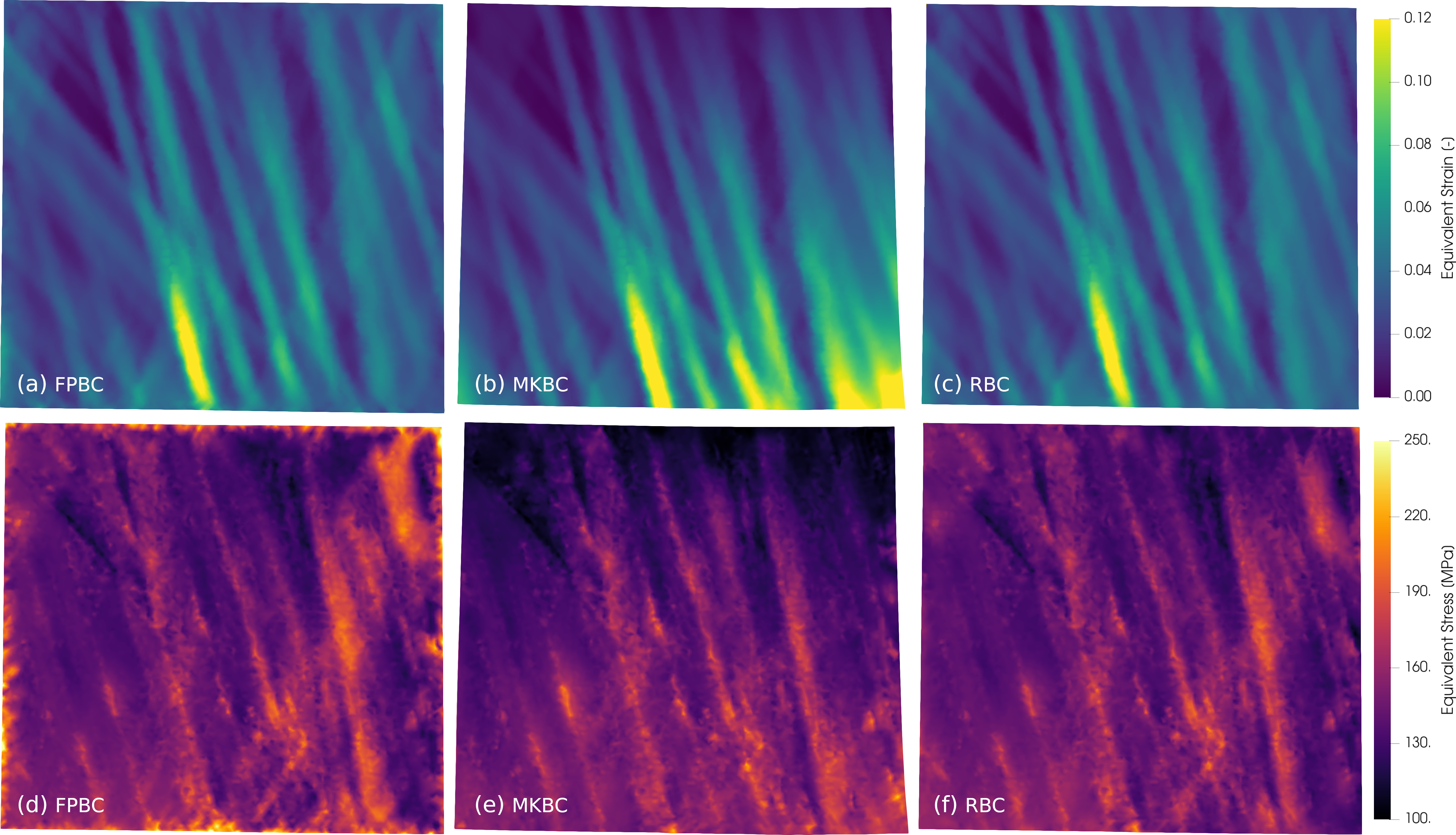}
		\phantomcaption \label{fig:bcs_results_a}
		\phantomcaption \label{fig:bcs_results_b}
		\phantomcaption \label{fig:bcs_results_c}
		\phantomcaption \label{fig:bcs_results_d}
		\phantomcaption \label{fig:bcs_results_e}
		\phantomcaption \label{fig:bcs_results_f}
	\end{subfigure}
	\caption{(a) Equivalent Green-Lagrange strain fields obtained in the ROI using the DSP model with (a) FPBC, (b) MKBC, and (c) RBC with $\tractionstiffness=10^5$ MPa. The equivalent Cauchy stress fields for the same simulations are shown in (d), (e), and (f), respectively.}
	\label{fig:bcs_results}
\end{figure}

FPBCs force the ROI to deform in the same way as in the experiment. However, this raises several concerns. For example, stress concentrations are introduced near the boundary because the material has to comply with the prescribed displacements, which in addition are affected by measurement noise. This can be observed in \Cref{fig:bcs_results_a,fig:bcs_results_d} that show the strain and stress field for the case in which the measured displacements on the boundary of the area in \Cref{fig:case0_overview} are directly applied to a simulation with the DSP model. In the equivalent stress field of \Cref{fig:bcs_results_d}, the highest stress concentrations occur close to the boundary. Measurement noise and errors also add to these stress concentrations. Even when these would be reduced by smoothing they may have a significant artificial detrimental effect, for example, when identifying material parameters with integrated digital image correlation \cite{rokos2018}. Another disadvantage is that these boundary conditions enforce strain localizations at measured locations of plastic slip crossing the boundary in the experimental specimen, while the (random) source distribution is generally different in the simulations and, therefore, triggers localizations elsewhere on the boundary. Close inspection of the equivalent strain field in \Cref{fig:bcs_results_a} shows that many localization bands inside the ROI are constrained close to the boundary.

To relax the FPBCs, the boundary $\boundary$ in the simulation is allowed to deviate from the measured displacements. This is done by introducing the following boundary condition:
\begin{equation}
	\traction = - \frac{\tractionstiffness}{L^*} \fluctuation \qquad \forall \vec{x} \in \boundary
	\ ,
	\label{eq:tractionpenalty}
\end{equation}
where $\tractionstiffness$ is a boundary stiffness, $L^*$ is a characteristic length, and $\fluctuation$ is the fluctuation vector given by
\begin{equation}
	\fluctuation = \dispvec - \dispvecmacro
	\label{eq:fluctuation}
	\ ,
\end{equation}
with $\dispvec$ the actual in-plane displacement in the simulation and $\dispvecmacro$ the measured displacement. \Cref{eq:tractionpenalty} applies a traction to the boundary of the volume element as a linear function of the difference between the actual and measured displacements. It forces the boundaries of the simulated volume element towards the measured displacements. Similar to the FPBCs, the $\dispvecmacro$ is extruded through the thickness of the boundary, such that $\traction$ is applied to the full boundary $\boundary$.

By only prescribing the traction at the boundaries, the average deformation in the region can deviate from the prescribed deformation. Therefore, the boundary condition of \Cref{eq:tractionpenalty} is complemented by a constraint that enforces the average deformation gradient in the simulation to be equal to the experimentally measured one. To connect the deformation in the 3D volume of a simulation to the 2D deformation of the surface area in the experiment, first, a mean in-plane displacement vector is defined by averaging through the thickness in the model:
\begin{equation}
	\dispvecmean = \frac{1}{L_z} \int_{z_1}^{z_2} \dispvec \, dz
	\ ,
	\label{eq:meandisp}
\end{equation}
where $L_z=z_2-z_1$ denotes the local thickness of the specimen. The area average of the in-plane deformation gradient calculated with the mean in-plane displacement vector is now enforced to take the same value as its experimental counterpart:
\begin{equation}
	\int_{\domaint}  \left( \nabla \dispvecmean \right)^T - \left( \nabla \dispvecmacro \right)^T \, d\domaint = \mathbf{0}
	\ ,
	\label{eq:volumeaveraging}
\end{equation}
where $\domaint$ represents the area of the ROI. Using the divergence theorem and consecutively substituting \Cref{eq:meandisp} and \Cref{eq:fluctuation} results in 
\begin{equation}
	\int_{\boundary} \frac{1}{L_z} \fluctuation \otimes \normal \, d\boundary = \mathbf{0}
	\ ,
	\label{eq:mkbc}
\end{equation}
where $\normal$ is the outward normal vector at the boundary. Since the constraint reduces to a boundary integral, only the displacements measured at the boundary of the simulation domain have to be used. Note that the resulting integral is computed on the 3D domain boundary $\boundary$. The term $1/L_z$ accounts for the non-uniform thickness of the specimen, i.e.\ it ensures that two opposing faces of a square plate with non-uniform thickness contribute equally to the constraint. For a constant thickness, this term can be taken outside of the integral and does not have an influence on the constraint.

The boundary conditions defined by \Cref{eq:tractionpenalty,eq:mkbc} are termed relaxed boundary conditions (RBCs) throughout the remainder of this paper. A similar set of boundary conditions was recently introduced by Wojciechowski \cite{wojciechowski2022} in the scope of computational homogenization. The amount of relaxation is determined by the stiffness $\tractionstiffness/L^*$ in \Cref{eq:tractionpenalty}. $L^*$ is adopted here as the length of the geometry in the direction perpendicular to the boundary. Since the considered regions have the same $x$ and $y$ dimensions, $L^*$ is the same for all boundaries $\Gamma$. In this way, $\fluctuation/L^*$ can be interpreted as the fluctuation in the (linear) average strain tensor over the domain contracted with the boundary normal, and $\tractionstiffness$ is the stiffness associated with that strain.

Two limit cases for $\tractionstiffness$ can be considered. For $\tractionstiffness \rightarrow \infty$, the fluctuation vanishes and the FPBCs are recovered. For the other limit case, $\tractionstiffness \rightarrow 0$, the constraint of \Cref{eq:tractionpenalty} vanishes, and only the average deformation is prescribed. This is similar to so-called minimal kinematic boundary conditions (MKBCs) used in computational homogenization frameworks, which often result in excessive localization near the weak spots at the boundary, while the rest of the domain stays undeformed \cite{mesarovic2005,inglis2008}. \Cref{fig:bcs_results_b,fig:bcs_results_e} show the same simulation as performed in \Cref{fig:bcs_results}, but with MKBCs. Localization is indeed more severe, e.g.\ in the bottom right corner, compared to the simulation with FPBCs, especially in the lower-right corner of the region.

\begin{figure}[!tb]
	\centering
	\includegraphics[width=.4\linewidth]{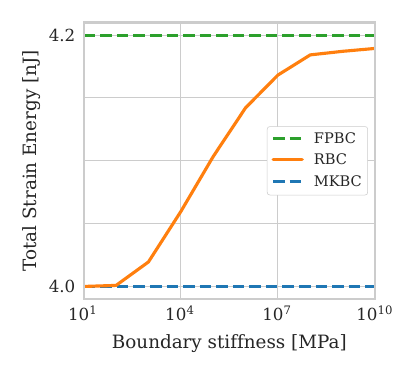}
	\caption{The total strain energy in the volume element as a function of the boundary stiffness $\tractionstiffness$ in the RBCs. The total strain energies obtained for FPBCs and MKBCs are denoted by dashed lines.}
	\label{fig:energy}
\end{figure}

Figure \ref{fig:energy} shows the total strain energy of the considered volume element for a range of traction stiffnesses for the simulation as done with the FPBCs and MKBCs of \Cref{fig:bcs_results}. The total strain energies obtained for the limit cases of MKBCs and FPBCs are also shown in the figure. Clearly, the total strain energy with RBCs changes from the value obtained with MKBCs to the value obtained with FPBCs with increasing $\tractionstiffness$. In this analysis, we adopt an intermediate traction stiffness of $\tractionstiffness=10^5$ MPa, which results in a total strain energy approximately halfway between the two limit cases. The equivalent strain and equivalent stress maps for this set of boundary conditions are shown in \Cref{fig:bcs_results_c,fig:bcs_results_f}. The deformations are similar to the results with FPBCs, but localizations near the boundary are less suppressed. Yet the unphysical strain peak in the bottom right corner of \Cref{fig:bcs_results_b} is now absent. Significant differences also appear in the stress field, where high values are no longer visible close to the boundary, in contrast to \Cref{fig:bcs_results_d}.

\subsection{Slip activity characteristics for comparison of experiments and simulations}
\label{section:localization}

The experimentally determined strain and slip fields are highly heterogeneous at the sub-grain level. The DSP model, unlike conventional CP, replicates this heterogeneity. However, the precise locations of slip bands and other features of the heterogeneity depend on the random properties of the slip system bands, i.e.\ simulated by the variability of the nucleation stress at the integration points, which represents the distribution of dislocation sources. This implies that each realization of the random microstructure will trigger different strain and slip activity patterns, none of which will be exactly identical to the experimental pattern. In addition, the width of the slip bands and other features in the modeling approach are set by the band width $\bandwidth$ introduced in the implementation of the DSP model, which, due to computational efficiency reasons, is typically larger than the width of the slip bands observed in the DIC strain fields. Note that the experimental strain patterns depend on the facet size used in the DIC analysis, which generally also broadens the true width of slip bands. These observations raise the question of how to objectively assess and compare the similarity (or dissimilarity) of both strain patterns with different length scales and configurations, which nevertheless may look similar to the observer's direct view.

\begin{figure}[!tb]
	\begin{subfigure}{\linewidth}
		\centering
		\includegraphics[width=.9\linewidth]{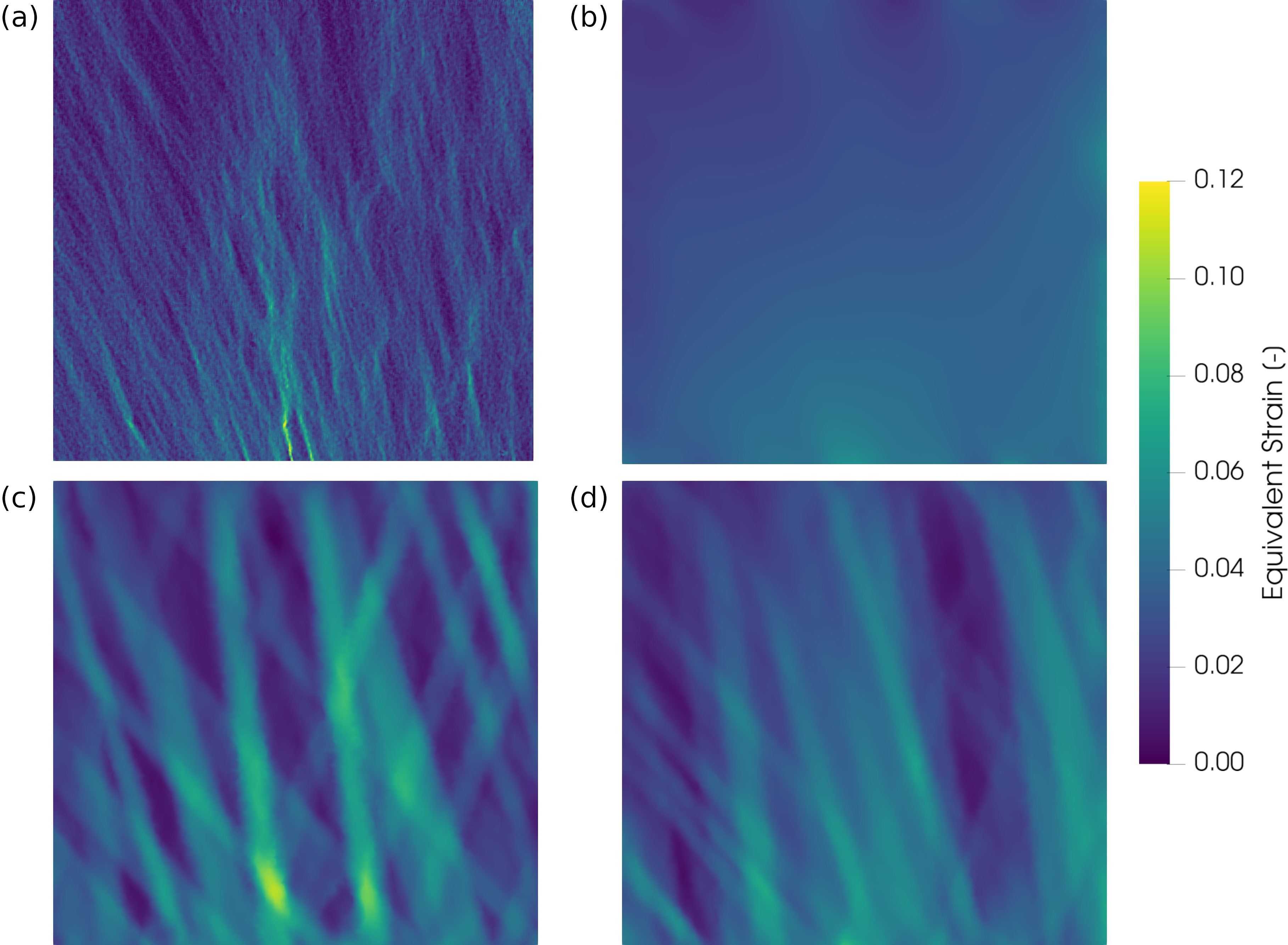}
		\phantomcaption \label{fig:case0_strainexp}
		\phantomcaption \label{fig:case0_straincp}
		\phantomcaption \label{fig:case0_strainbest}
		\phantomcaption \label{fig:case0_strainworst}
	\end{subfigure}
	\caption{The 2D equivalent Green-Lagrange strain field (a) in the experiment, (b) obtained with the CP model and obtained for the (c) "best" and (d) "worst" realizations with the DSP model.}
	\label{fig:case0_strain}
\end{figure}

Before addressing this question, we first illustrate the problem for the single-grain ferrite region considered above. In both the CP simulation and the DSP simulations, 12 $\slipfamilyA$ and 12 $\slipfamilyB$ slip systems are taken into account. Since the out-of-plane deformation in the experiment is unknown, a 2D equivalent strain measure based on the in-plane components of the Green-Lagrange strain tensor is used for plotting the deformation in both the experiment and the simulations \cite{Tasan2014}: $E_\text{eq} = \tfrac{\sqrt{2}}{3} \sqrt{\left(E_{xx}-E_{yy}\right)^2 +E_{xx}^2+E_{yy}^2 + 6E_{xy}^2}$. The resulting strain fields are shown in \Cref{fig:case0_strain}, where \Cref{fig:case0_strainexp} displays the experimental equivalent strain field and \Cref{fig:case0_strainbest,fig:case0_strainworst} the computational strain fields for two realizations of the random source distribution. These two realizations are termed "best" and "worst", which will be clarified later in this section. The result obtained with conventional CP is also shown, for reference, in \Cref{fig:case0_straincp}. In the latter, the deformation is nearly homogeneous. Only the strains around the bottom center of the region are somewhat higher due to the thickness profile and the boundary conditions. The experimental data, in \Cref{fig:case0_strainexp}, on the other hand, shows a fine-scale pattern of intense localization bands, in which the strain locally exceeds the mean value by a factor of 10 or more. Clearly, the maximum strain in the region is much higher in the experiment than in the CP result. 

\begin{figure}[!tb]
	\begin{subfigure}{\linewidth}
		\centering
		\includegraphics[width=.9\linewidth]{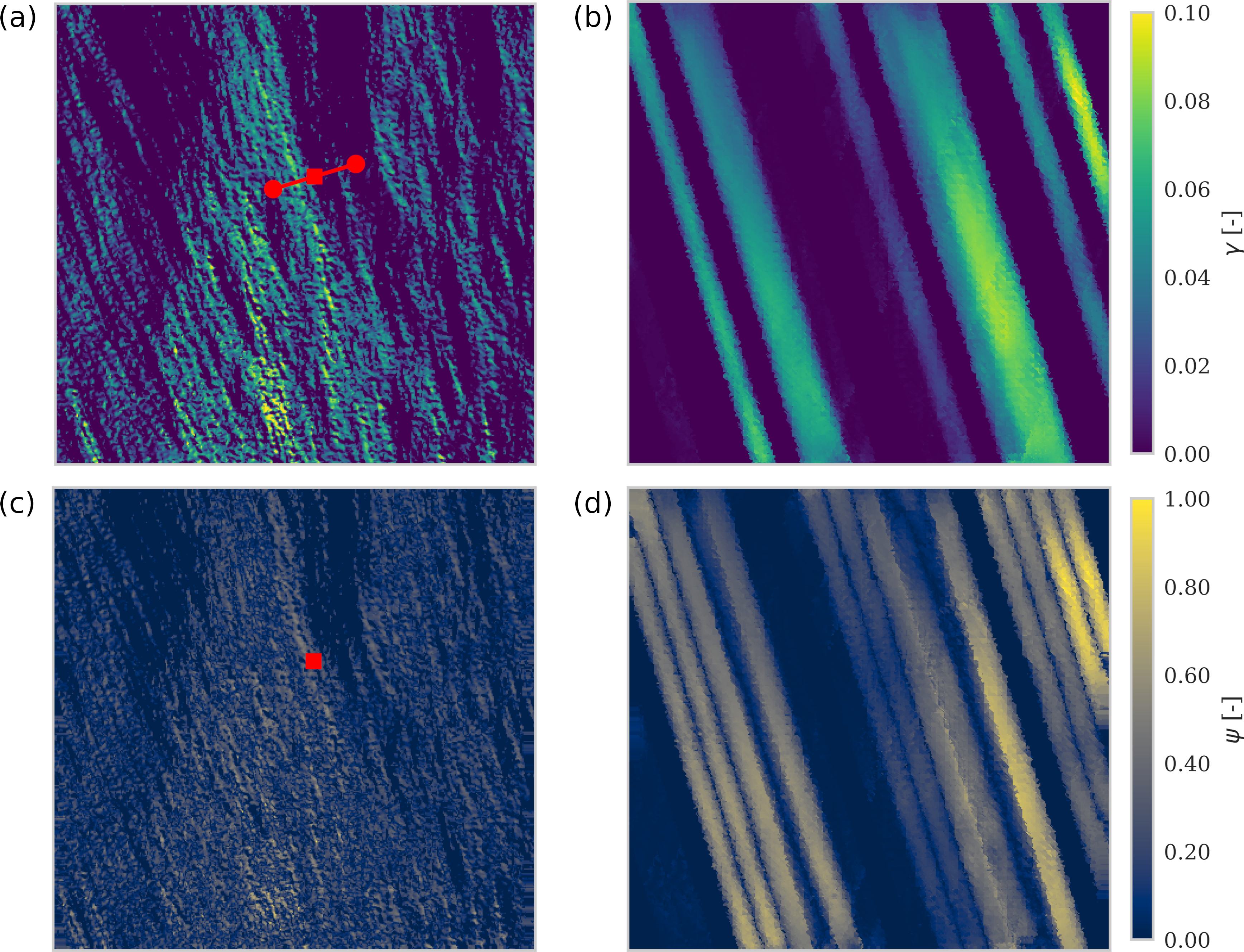}
		\phantomcaption \label{fig:case0_exp6_slip}
		\phantomcaption \label{fig:case0_dsp6_slip}
		\phantomcaption \label{fig:case0_exp6_loc}
		\phantomcaption \label{fig:case0_dsp6_loc}
	\end{subfigure}
	\caption{The slip activity fields of the most active slip system, i.e.\ slip system $(110)[\bar{1}11]$ marked by SS1 in the pole figure of \Cref{fig:case0_overview_d}, for (a) the experiment of \Cref{fig:case0_strainexp} and (b) the DSP simulation of \Cref{fig:case0_strainbest}. The localization quantity field of SS1 in (c) the experiment and (d) the DSP simulation.}
	\label{fig:case0_exp6}
\end{figure}

The equivalent strain fields for the two different realizations of the DSP model reveal fluctuations that are at least qualitatively similar to those observed in the experiment. In particular, the predicted bands have similar orientations to those in the experimental result, and their amplitude peaks in the same region near the bottom of the considered ROI. However, as expected, neither the width of the bands nor their precise location matches the experiment. Similar observations may be made for the slip distributions associated with the individual slip systems. For the modeling, this data is readily obtained from the simulations. For the experiments, it may be extracted from the measured deformation fields by the SSLIP method, introduced in Vermeij et al.\ \cite{Vermeij2023}. This method locally decomposes the deformation gradient field, measured on the specimen surface, into contributions from the theoretical slip systems. This is achieved by solving an optimization problem in which the measured kinematics optimally captures a combination of slip systems, each with a to-be-determined amount of slip. As an example, the slip activity of the most active slip system in the experiment denoted as SS1 in the pole figure of \Cref{fig:case0_overview_d} is shown in \Cref{fig:case0_exp6_slip}. The slip activity of the same slip system in the DSP simulation of \Cref{fig:case0_strainbest} can be seen in \Cref{fig:case0_dsp6_slip}.

Obtaining a perfect match between the slip activity and, consequently, the equivalent strain map in a simulation and in the experiment is not possible. This is because, at the considered scale, stochastic fluctuations have a prominent effect on the experiment. These stochastic fluctuations are also present in the DSP model. Therefore, a comparison is made based on the amount of slip on the individual slip systems and the degree of localization in the slip fields. However, the spatial localizations observed in both the DSP model as well as the experiments are limited due to numerical and experimental resolutions. Accordingly, a measure is introduced that represents the amount of localization, given a certain length scale.

A typical characteristic of strong localizations is a high gradient in the strain field. Localization of a specific slip system usually follows its slip plane trace. Therefore, the gradient of the slip activity field perpendicular to the slip plane trace is a measure of fluctuations in that slip system around a particular point. However, the fluctuations in the deformation field obtained with the DSP model are dependent on the band width $l$. Preferably, the adopted band width is as small as possible. It is, however, limited due to the computational cost of a simulation. Therefore, a localization quantity, $\locvalue$, is introduced that approximates the gradient in the slip activity field by the difference in slip activity of two points separated by a vector $\locdistance$ (similar to a central difference approximation). This vector is perpendicular to the in-plane trace of the considered slip plane and has a length, $\locdistance$, that is equal to the adopted band width in the DSP model, projected on the $xy$-plane, since this is the smallest length over which fluctuations in the simulations can be resolved. Note that the smallest length over which fluctuations can take place in the experimental data is also limited, for example, by the subset size used in DIC. However, for comparing the two data sets the largest of these length scales should be used. i.e.\ the numerical band width.

The localization quantity is given by
\begin{equation}
	\locvalue(\coordvec) = \frac{ | |\shearstrain ( \coordvec - \tfrac{1}{2}\locdistance )| - |\shearstrain ( \coordvec + \tfrac{1}{2}\locdistance )| | }{ |\max_{\coordvec} \left( \shearstrain \right)| }
	\ ,
	\label{eq:locvalue}
\end{equation}
where $\shearstrain$ is the slip activity. It is normalized by the maximum value of the slip activity in the specimen. In this way, $\locvalue$ takes a value of 1 when two points separated by $\locdistance$ have slip activity equal to 0 and $\max_{\coordvec} \left( \shearstrain \right)$.

The localization quantity of the slip activity field shown in \Cref{fig:case0_exp6_slip} is calculated by Equation \ref{eq:locvalue}. This is schematically depicted by the red line. The localization value at the midpoint of this line, i.e.\ the red square, is calculated by the slip activity at the end of this line, i.e.\ the red dots. A localization quantity field is obtained by moving the midpoint of this red line over all the pixels of the slip activity field. The result is shown in \Cref{fig:case0_exp6_loc}. Note that the length of the red line in \Cref{fig:case0_exp6_slip} is not the actual length of $\locdistance$ that is used in the analysis, but is magnified for clarification. The localization quantity field calculated from the slip activity field of \Cref{fig:case0_dsp6_slip} is displayed in \Cref{fig:case0_dsp6_loc}.

\begin{figure}[!tb]
	\centering
	\begin{subfigure}{\linewidth}
		\includegraphics[width=\linewidth]{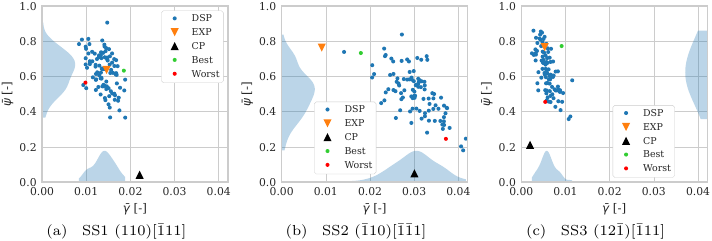}
		\phantomcaption	\label{fig:realizations1}
		\phantomcaption \label{fig:realizations2}
		\phantomcaption \label{fig:realizations3}
	\end{subfigure}
	\caption{The average slip activity and localization value of the region of \Cref{fig:case0_overview} for the experiment, 100 DSP simulations, and the CP simulation, for three different slip systems. The kernel density estimates of the DSP simulations is plotted on both axes.}
	\label{fig:realizations}
\end{figure}

The slip activity and localization fields are defined locally. However, in order to examine many results together, both measures were averaged over the region. Figure \ref{fig:realizations} displays the resulting values for the experiment, the CP model, and 100 realizations of the DSP model for the three most active slip systems in the experiment. The horizontal and vertical axes represent the average slip activity, $\bar{\shearstrain}$, and the average localization quantity, $\bar{\locvalue}$, respectively. 

For the most active slip system in the experiment, i.e.\ the $(110)[\bar111]$ slip system denoted by SS1, shown in Figure \ref{fig:realizations1}, the experimental data point lies in the center of the point cloud of the DSP results, close to the mean. This indicates that the experiment, at least in terms of the two characteristics employed here, may be captured as a particular realization of the ensemble of DSP models represented by the blue point cloud. The CP simulation has a higher average slip activity than the experiment and DSP simulations. Furthermore, the localization quantity for the CP simulation is much lower, which can be expected based on the strain field in \Cref{fig:case0_straincp}.

The results for the second most active slip system in the experiment, i.e.\ the $(\bar110)[\bar1\bar11]$ slip system denoted by SS2, are shown in Figure \ref{fig:realizations2}. The amount of scatter in the DSP results, particularly in the average slip activity $\bar{\gamma}$, is larger than in \Cref{fig:realizations1}. Nevertheless, the average slip activity in the experiment is lower than in all the simulations. The average slip activities for the DSP simulations are located around the average slip activity obtained with the CP model. The localization value of the experiment is similar to a few DSP results, but most simulations have a lower value.

Figure \ref{fig:realizations3} shows the results for the third most active slip system in the experiments (SS3), i.e.\ the $(12\bar1)[\bar111]$ slip system denoted by SS3. The experimental data point lies in the point cloud of the DSP results. Contrary to the most active slip system, the slip activity in the CP simulation is lower than in the experiment and the DSP simulations.

For most slip system specific quantities, the experiment falls within the range of the DSP simulations. Only SS2 is significantly more active in most DSP simulations. Interestingly, there is no slip system compensating for this higher activity, i.e.\ there is no slip system that is significantly more active in the experiment than in the simulations. A possible explanation is that, in contrast to the in-plane deformation, the average out-of-plane deformation in the simulations is not enforced to be equal to that of the experiment. The difference in average slip activity between the CP simulation and the DSP simulations for SS1 and SS2 shows that introducing fluctuations in the slip resistance not only results in a more localized strain field but also influences the relative activity of the slip systems. In general, a better match between experiment and simulation is obtained with the DSP model compared to the CP model, for both the slip fluctuations as well as the slip activity. Our previous study \cite{Wijnen2023} showed that the DSP model gives a more accurate slip system activity compared to the CP model for uniaxial tensile tests. The results presented in this section show that this conclusion can be extended to other geometries with more complex loading conditions.

It is instructive to consider, in the results of \Cref{fig:realizations}, those DSP simulations that match the experiment the best and worst, solely in terms of $\bar{\gamma}$ and $\psi$. The absolute difference between the experiment and all the simulations is calculated, for both the slip activity and localization value. The differences are normalized by the maximum difference between the experiment and a simulation, such that for both properties there is a single data point with a value of 1 for every slip system. Next, the distance between the experiment and a simulation for a single slip system is calculated by taking the Euclidean norm of the vector that contains the normalized difference in slip activity and localization value. Finally, the distances of the three slip systems are added up, resulting in a single distance per DSP simulation. The simulation that matches "best" (shortest distance) and "worst" (largest distance) with the experiment are both marked in the plots of Figure \ref{fig:realizations}. It can be observed that the best simulation has a localization value that is similar to the experiment for all three slip systems. The slip magnitude for SS2 of the best simulation is relatively low, closer to the experiment than most other DSP simulations. To compensate for the low slip magnitude of SS2, the slip magnitudes of SS1 and SS3 are relatively high.

The equivalent strain fields of the best and worst matching simulations are shown in \Cref{fig:case0_strainbest,fig:case0_strainworst}, respectively. These figures show that the worst matching simulation has less pronounced strain fluctuations than the best matching simulation. The latter contains more small areas with an equivalent strain close to zero and higher equivalent strain values in general. The smoother strain field of the worst agreeing simulation can also be recognized in the data of \Cref{fig:realizations}, where it reveals a low localization value for all three slip systems.


\section{Case study of ROI containing a grain boundary}
\label{section:case1}

\begin{figure}[!htb]
	\centering
	\begin{subfigure}{\linewidth}
		\centering
		\includegraphics[width=\linewidth]{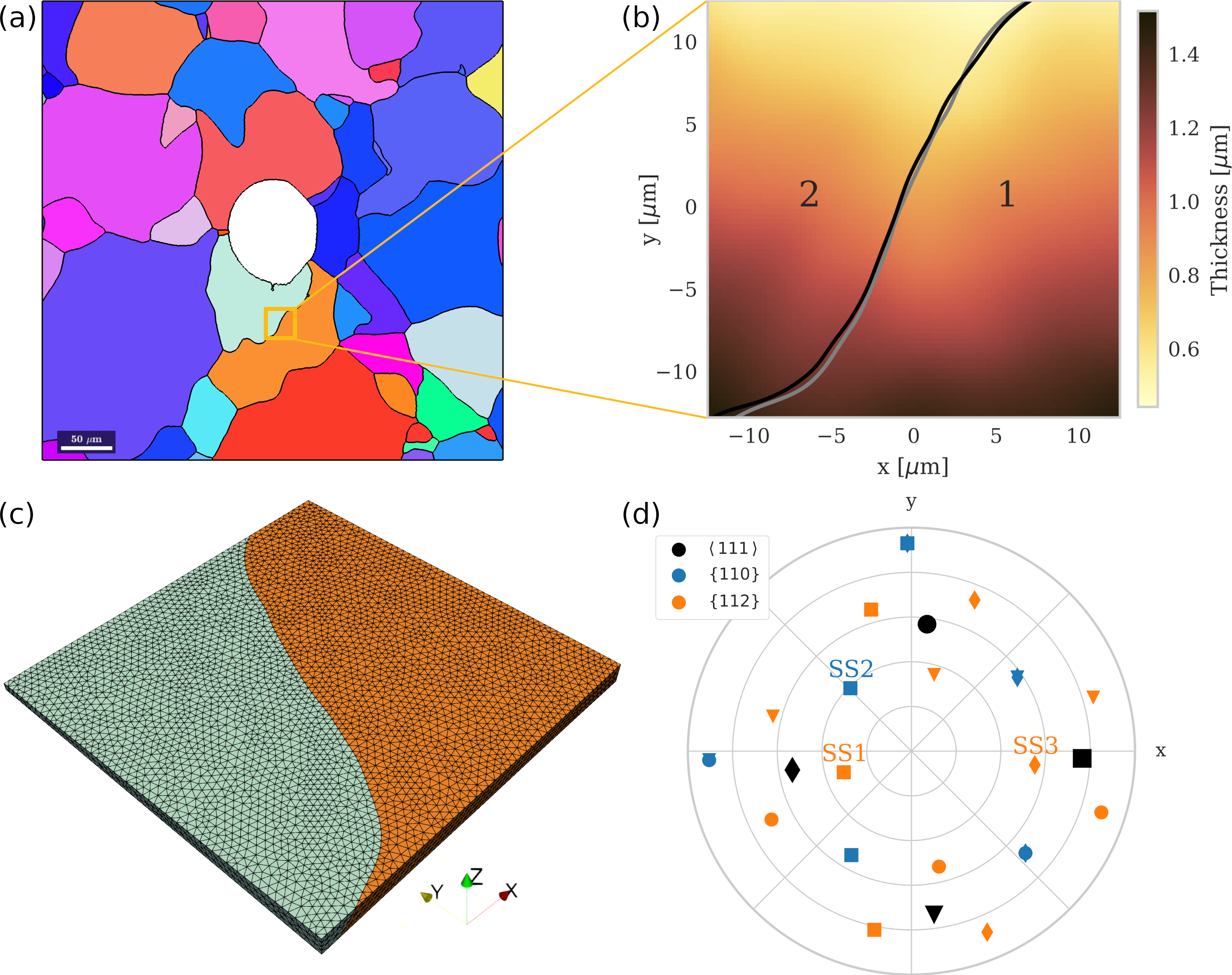}
		\phantomcaption \label{fig:case1_overview_a}
		\phantomcaption \label{fig:case1_overview_b}
		\phantomcaption \label{fig:case1_overview_c}
		\phantomcaption \label{fig:case1_overview_d}
	\end{subfigure}
	\caption{(a) EBSD map of the indicated area with a grain boundary around the hole of the specimen. The smaller ROI is marked with a yellow rectangle. (b) The thickness profile of the ROI, where the grain boundary edges are shown as an overlay with a black line for the front surface and a gray line for the rear surface. (c) Discretization of the ROI consisting of 64,366 quadratic tetrahedral elements and 103,136 nodes. (d) The orientations of the slip plane normals and slip directions of the grain 1. The blue and orange markers denote the ${110}$ and ${112}$ plane normal, respectively, while the slip directions are denoted by black markers. The plane normal of a slip system is depicted with the same marker as its slip direction.}
	\label{fig:case1_overview}
\end{figure}

So far, the framework has been applied to a relatively simple ROI, i.e.\ a single-phase region without grain boundaries. In this section, a ferrite region containing two grains is examined. The ROI is taken from the same specimen as the ROI considered in \Cref{section:methodology}. Its location is marked in \Cref{fig:case1_overview_a}. The thickness profile of the specimen, together with the grain boundaries on the front surface (black line) and rear surface (gray line), are depicted in \Cref{fig:case1_overview_b}. The region is approximately split in half by a curved grain boundary. To perform simulations, the 3D geometry was discretized with 64,366 quadratic tetrahedral elements and 103,136 nodes, as depicted in \Cref{fig:case1_overview_c}. The pole figure in \Cref{fig:case1_overview_d} shows the orientation of the slip systems in grain 1, denoted in \Cref{fig:case1_overview_b}.

\subsection{Equivalent strain fields}

\begin{figure}[!tb]
	\begin{subfigure}{\linewidth}
		\centering
		\includegraphics[width=.9\linewidth]{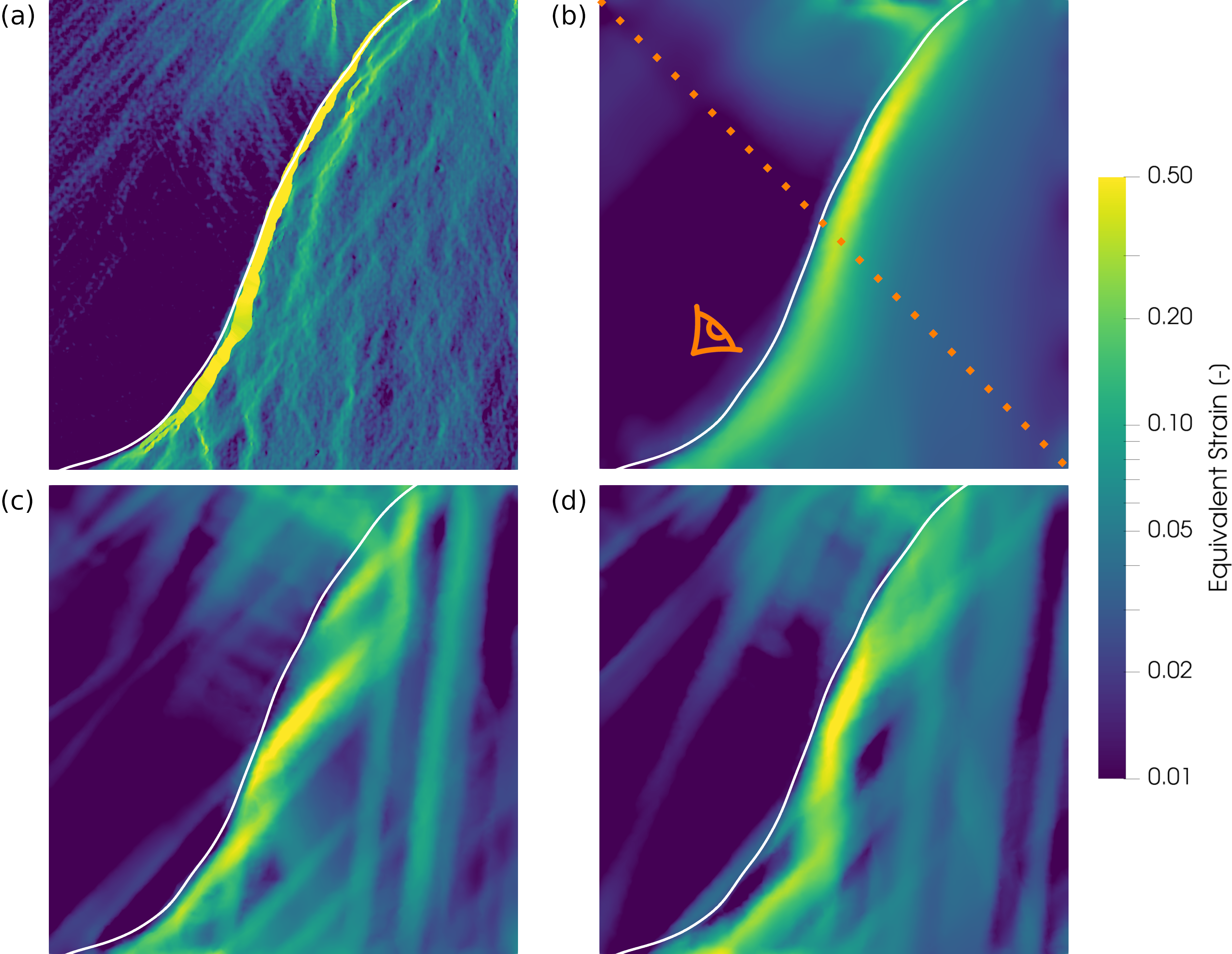}
		\phantomcaption \label{fig:case1_strainexp}
		\phantomcaption \label{fig:case1_straincp}
		\phantomcaption \label{fig:case1_strainbest}
		\phantomcaption \label{fig:case1_strainworst}
	\end{subfigure}
	\caption{2D equivalent Green-Lagrange strain fields of the ROI shown in \Cref{fig:case1_overview} for (a) the experiment, (b) the CP simulation, (c) the "best" DSP simulation and (d)  the "worst" DSP simulation. The orange dotted line and the eye in (b) show the location of the cross-section and camera angle that will be used later in \Cref{fig:case1_3D}.}
	\label{fig:case1_strain}
\end{figure}

The 2D equivalent strain fields of the experiment are presented in \Cref{fig:case1_strainexp}. At some positions, predominantly near the grain boundary, DIC data was not correctly correlated due to large deformations. Therefore, interpolation was performed on the displacement fields to recover the complete deformation fields, as discussed in Vermeij et al.\ \cite{Vermeij2023b}. The equivalent strain field shows a strong localization close to the boundary. Yet, a detailed inspection of the secondary electron (SE) and backscatter electron (BSE) images did not reveal clear signs of any damage. Furthermore, due to the high resolution and the precise alignment of the images, the strong localization was identified to be located in grain 1, i.e.\ at the right side of the grain boundary. No indication of grain boundary sliding was observed. Away from the boundary, grain 1 shows many fluctuations in the strain fields. Grain 2, i.e.\ the left grain, shows only small strain localizations coming in at the top of the region, propagating through approximately one-third of the grain.

\Cref{fig:case1_straincp} shows the equivalent strain field obtained with the conventional CP model. It reveals high strains in grain 1 near the grain boundary, similar to the experiment. The strain gradually decreases further away from the boundary. Clearly, the strain fields are again much smoother compared to the experiment.

The equivalent strain fields for the "worst" and "best" realizations obtained with the DSP model are shown in \Cref{fig:case1_strainbest,fig:case1_strainworst}, respectively. The simulations show more fluctuations in the strain field than the conventional CP simulations. However, in both simulations, strain bands tend to line up with the grain boundary in grain 1. This indicates that the influence of fluctuations on the slip resistance is less pronounced close to the grain boundary. Instead, the morphology of the grain boundary dictates the degree of plasticity here. In grain 2, similar strain localizations as in the experiments are observed, especially for the simulation of \Cref{fig:case1_strainbest}. Compared to the CP result, the strains in grain 2 obtained with the DSP model propagate further downwards.

\subsection{Statistical slip system analysis}

\begin{figure}[!tb]
	\centering 
	\begin{subfigure}{\linewidth}
		\includegraphics[width=\linewidth]{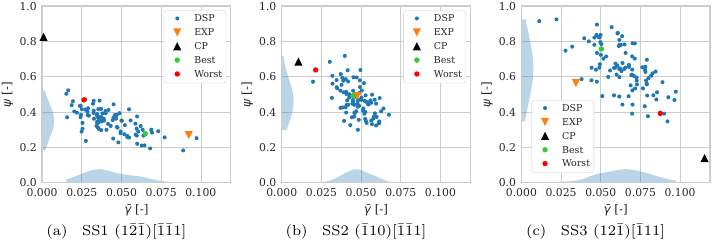}
		\phantomcaption	\label{fig:case1_realizations1} 
		\phantomcaption \label{fig:case1_realizations2} 
		\phantomcaption \label{fig:case1_realizations3}
	\end{subfigure}
	\caption{The average slip magnitude and localization value in grain 1 of \Cref{fig:case1_overview} for the experiment, 100 DSP simulations, and the CP simulation, for three different slip systems. The kernel density estimates of the DSP simulations is plotted on both axes.}
	\label{fig:case1_realizations}
\end{figure}

Despite the lack of fluctuations in the strain field, the conventional CP model appears to describe the deformation in the experiment reasonably well in an average sense. However, a more detailed analysis with a focus on slip system activities and degree of localization, similar to \Cref{section:localization}, was performed here as well. Since the majority of the deformation took place in grain 1, the analysis is restricted to this grain. The strain fields are again decomposed into the contributions of the individual slip systems, both experimentally and numerically, and the localization quantity is calculated through \Cref{eq:locvalue}. The averages of the slip activity and the localization quantity in grain 1 in the experiment, the conventional CP simulation, and 100 DSP simulations are shown in \Cref{fig:case1_realizations}, for the three most active slip systems in the experiment. These are the $(1\bar21)[\bar1\bar11]$, $(110)[\bar1\bar11]$ and $(12\bar1)[\bar111]$ slip systems, denoted by SS1, SS2 and SS3, respectively. They are ordered based on their average slip activity in the experiment. Their orientations are indicated in the pole figure of \Cref{fig:case1_overview_d}.

Surprisingly, the most active slip system in the experiment (SS1) is not at all active in the conventional CP simulation. Instead, most slip in the CP simulation emerges from SS3 (\Cref{fig:case1_realizations2}), where the slip magnitude is significantly larger than in the experiment. Although the ROI is deformed via a complex loading path, its proximity the the hole in the sample renders the deformation in the x-direction dominant. The Schmid factor of SS3 for uniaxial tension in the $x$-direction is close to 0.5, which means that this slip system is likely to activate. However, due to the microstructure of this region, strain bands aligned with the grain boundary tend to form. In the conventional CP simulation, this strain band is almost fully accommodated through slip on SS3, even when its slip plane is not fully aligned with the grain boundary, by a combination of slip and kinking. The kink mechanism is a localization mode inherent to the constitutive behavior of crystal plasticity, in which the localization band is oriented perpendicular to the slip plane  \cite{Asaro1977,Forest1998}. Physically, kinking requires a large number of dislocation sources that are aligned perpendicular to the slip plane \cite{marano2019}. Furthermore, it introduces a high geometrically necessary dislocation (GND) density which is energetically unfavorable \cite{marano2021}. Therefore, in the experiment, deformation modes parallel to the slip systems are preferred and SS3 is not activated. Instead, the strain band following the grain boundary is a result of slip on both SS1 and SS3. Because these slip systems have the same slip direction, cross-slip can occur between them, as confirmed by the experimental strain map in which the localization band is curving along the grain boundary (\Cref{fig:case1_strain}a).

In the DSP simulations, slip does take place on SS1, as well as on SS3. For most simulations, the average slip activity on SS1 is lower than in the experiment, but several realizations have a similar average slip activity. Also, the slip activities of SS2 and SS3 in the DSP simulations are similar to those in the experiment. However, the amount of slip on SS3 is slightly overpredicted in most realizations. Because the slip resistance of a slip system in the DSP model varies in the direction perpendicular to the slip plane due to the limited number of dislocation sources, the slip mechanism parallel to the slip planes is much more favorable compared to the kink mechanism, as is the case in the experiment. As a result, the activation of SS3 in the strain band near the grain boundary is suppressed. \Cref{fig:case1_strainbest,fig:case1_strainworst} reveal that the strain band following the grain boundary consists of short slip bands on alternating slip systems, resulting in a zig-zag pattern. This alternation happens mainly between SS1 and SS2. Although no cross-slip mechanism is incorporated explicitly into the model, the two slip systems having the same slip direction are most likely preferred due to compatibility. Note that a combination of the slip planes of SS1 and SS3 also results in a direction approximately parallel to the grain boundary, but that out-of-plane components of these slip directions are significantly different, entailing deformation incompatibilities.

The average localization quantity in grain 1 of the experiment lies in the point cloud of the DSP simulations for all three slip systems. The "best" and "worst" matching DSP simulations with the experiment were again determined based on the average quantities, as described before. The equivalent strain fields of the "best" and "worst" matching simulations (green and red dots in \Cref{fig:case1_realizations}, respectively) are shown in \Cref{fig:case1_strainbest,fig:case1_strainworst}, respectively. When only considering these equivalent strain fields, \Cref{fig:case1_strainbest} does not seem to be much closer to the experiment than \Cref{fig:case1_strainworst}. However, \Cref{fig:case1_realizations} reveals that the ratios of slip system activities of the "best" matching DSP simulation are much closer to the experiment than the "worst" matching DSP simulation. 

\subsection{Slip system fields}

\begin{figure}[!tb]
	\begin{subfigure}{\linewidth}
		\centering
		\includegraphics[width=\linewidth]{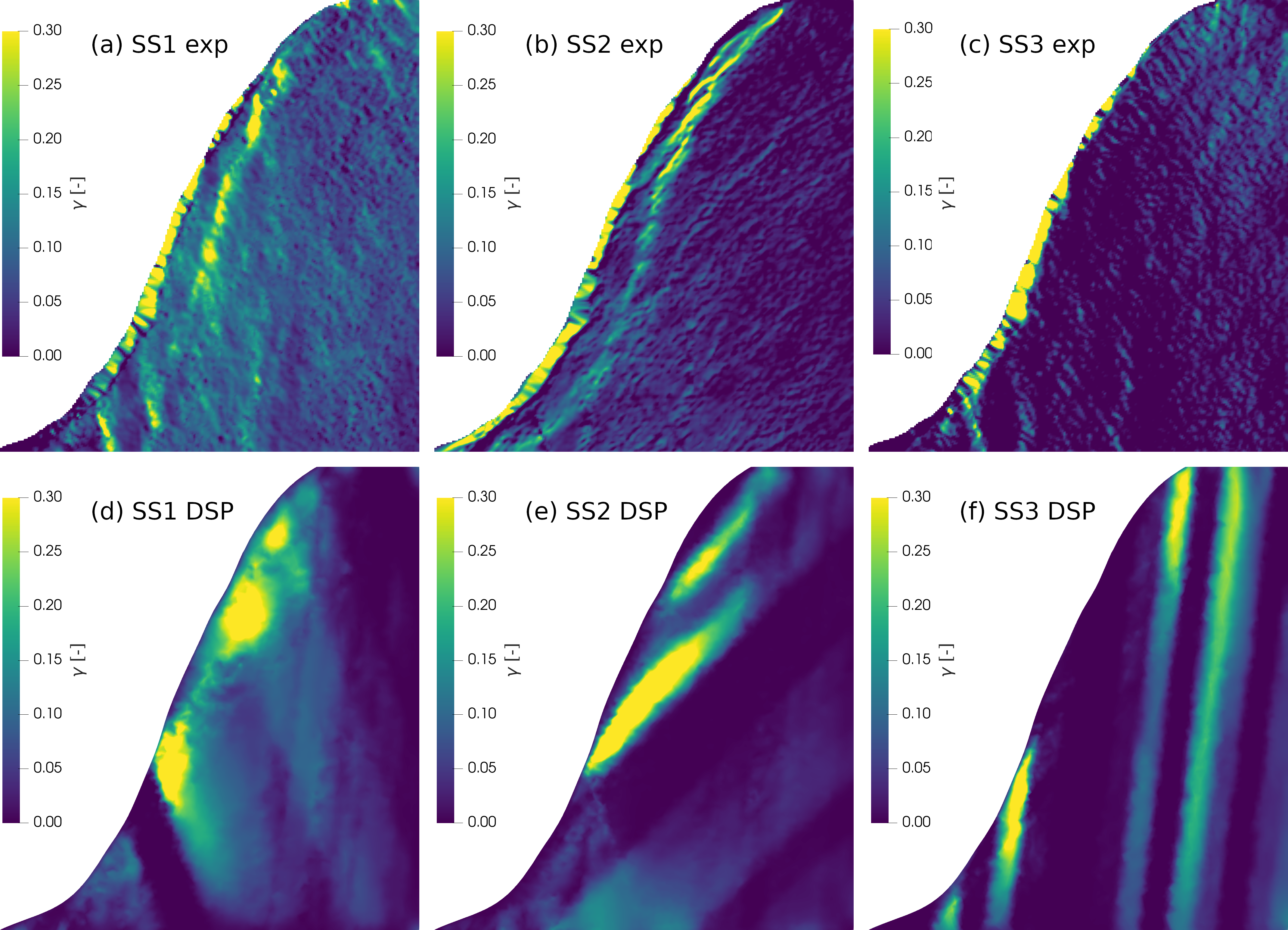}
		\phantomcaption \label{fig:case1_slipsystems_a}
		\phantomcaption \label{fig:case1_slipsystems_b}
		\phantomcaption \label{fig:case1_slipsystems_c}
		\phantomcaption \label{fig:case1_slipsystems_d}
		\phantomcaption \label{fig:case1_slipsystems_e}
		\phantomcaption \label{fig:case1_slipsystems_f}
	\end{subfigure}
	\caption{(a-c) Slip activity fields of the three most active slip systems in the experiment and (d-f) the slip activity fields of the corresponding slip systems in the "best" matching DSP realization.} 
	\label{fig:case1_slipsystems}
\end{figure}

The "best" matching DSP simulation is most similar to the experiment in terms of the average characteristics of the slip fields. However, this provides no information on the spatial agreement between slip fields. \Cref{fig:case1_slipsystems} presents the slip activity fields of the three most active slip systems in grain 1 for the experiment in (a-c) and for the "best" realization of the DSP model in (e-f). Near the grain boundary, all three slip systems reveal a significant amount of slip in the experiment. Furthermore, SS1 is active throughout the entire grain. The activity of SS2 is most pronounced in the region close to the grain boundary. Slip on SS3 is also observed in the upper-right region of the grain.

In the DSP simulation, SS1 is most active near the grain boundary, while slip is observed throughout almost the entire grain, similar to the experiment. Localization bands of SS2 are visible close to the grain boundary. The amount of slip in these localization bands fades out further away from the grain boundary. SS3 has slip bands visible in the right part of the grain, primarily in the upper-right region. Furthermore, a small slip band is observed in the lower-left region, against the grain boundary. So it can be concluded that regions of the grain in which the particular slip systems are active, are in adequate agreement between the experiment and the "best" DSP simulation. Note again that a perfect match is not feasible since both the experiment and the DSP simulation are significantly influenced by the stochastic fluctuations. Furthermore, this DSP realization closest resembles the experiment based on the average characteristics of the slip fields, and not on the exact position of the localizations or the width of the slip bands.

\subsection{3D deformation}

\begin{figure}[!tb]
	\centering
	\includegraphics[width=.8\linewidth]{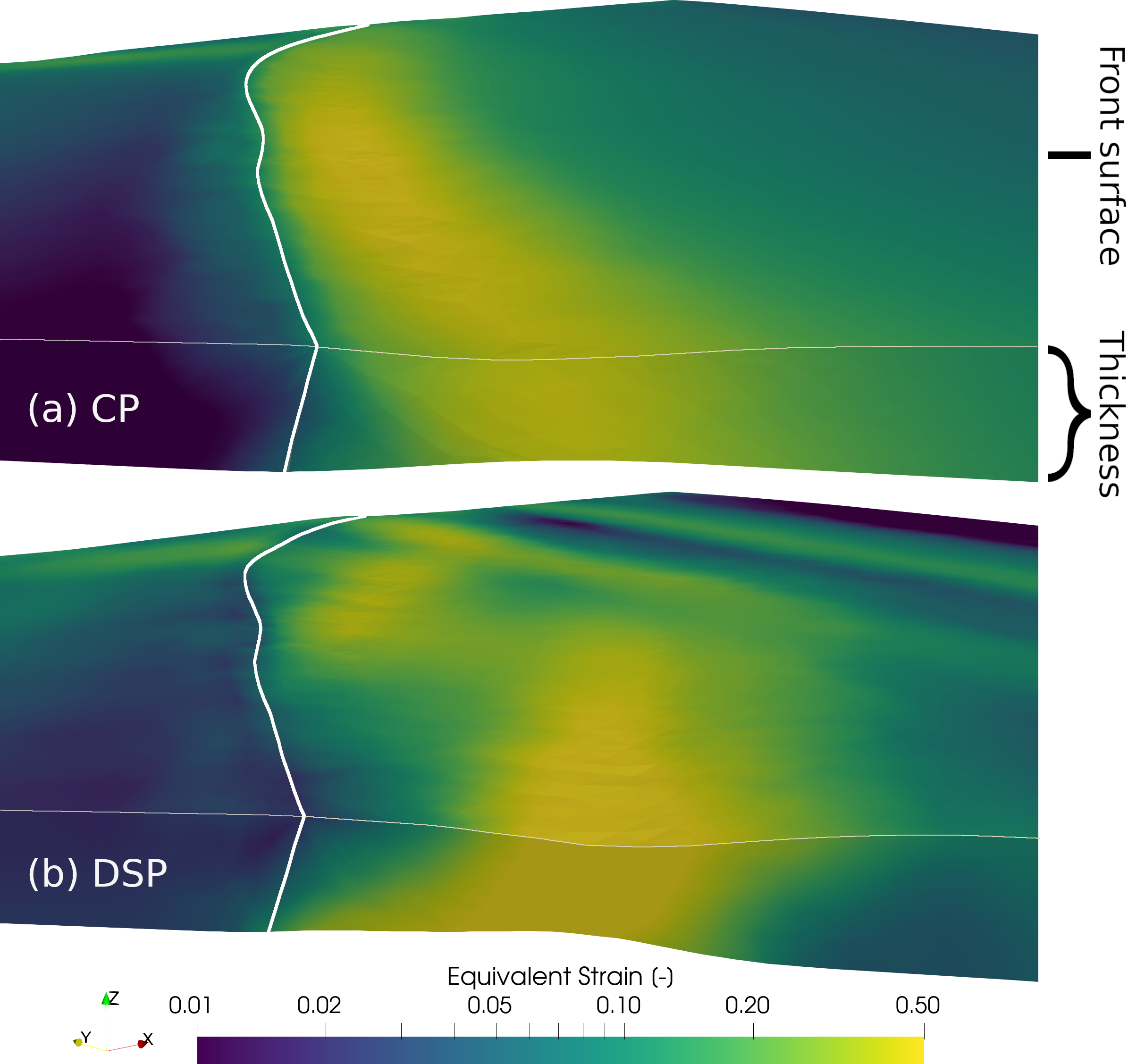}
	\caption{3D cross-sections of the deformed geometries for (a) the conventional CP model and (b) the "best" realization with the DSP model. The deformation is on the true scale. The colormap shows the (3D) equivalent Green-Lagrange strain. The cross-sections are taken from the upper-left corner to the lower-right corner of the region. The viewpoint is approximately from the lower-right corner, looking at the front surface.} 
	\label{fig:case1_3D}
\end{figure}

Contrary to the experimental data, the numerical data is fully 3D. This allows us to inspect the 3D through-thickness deformation of the considered region. \Cref{fig:case1_3D} shows the deformed cross-sections of the conventional CP simulation in subfigure (a) and of the "best" DSP simulation in subfigure (b), from a perspective view. The cross-sections were taken from the upper-left corner to the lower-right corner of the regions while the viewpoint is from approximately the lower-left corner, looking at the front surface, as depicted by the orange dotted line and eye symbol in \Cref{fig:case1_overview_b}.

Despite the 2D character of the ultra-thin geometry, the equivalent strain fields clearly reveal the 3D orientation of the slip systems. In the DSP simulation, a strain band is observed that touches the grain boundary on the rear. The strain band follows the orientation of the slip plane of SS2 (see \Cref{fig:case1_overview_d}). As a result, the same strain band observed on the front is located at a small distance from the grain boundary. This shift of the strain localization with respect to the front and rear is much less pronounced in the CP model. Here, the strain band is the result of slip on SS3. The slip plane of this slip system makes an angle of approximately 45$\degree$ with the xy-plane, i.e.\ the plane of the geometry, however, the strain localization is roughly straight through the thickness of the geometry. This again demonstrates that the conventional CP model has no restrictions in forming strain bands that are not aligned to the slip plane of the active slip system, in contrast to commonly observed slip system mechanics.


\section{Summary and conclusions}
\label{section:conclusions}

In this study, a detailed numerical-experimental analysis was performed on thin regions of a ferrite polycrystalline sheet. The quasi-2D nature of the microstructure enabled us to fully characterize the specimen's morphology and deformation. This enables direct comparisons between experiments and simulations. The richness of the available experimental data allows to assess and possibly improve current materials models to include more relevant physical phenomena.

The framework was used to study the deformation kinematics of two ferrite regions. This was done at the level of individual slip systems. Two types of models were employed, namely, a conventional crystal plasticity model, in which the average behavior of many underlying atomic planes is accounted for, and the recently introduced discrete slip plane (DSP) model, which considers fluctuations in slip resistances on atomic planes due to stochastic variability in the location and strength of dislocation sources.

A detailed analysis of 100 DSP realizations revealed a remarkable difference in slip system activity between conventional CP simulations on the one hand and DSP simulations and the experiment on the other hand, especially for the ferrite region considered in the case study of \Cref{section:case1}. Although the deformation fields obtained with the conventional homogeneous CP model may appear in reasonable agreement with the experiment at first sight, the conventional CP model completely failed to predict the correct active slip systems. Almost all deformation took place on one favorably oriented slip system, following a strain path that was not perpendicular to the slip system. In contrast, the fluctuations introduced by the DSP model through the stochastics of dislocation sources facilitate strain bands parallel to the slip systems. As a result, the grain boundary in the considered region had a significant effect on the slip system activity in the DSP simulations, similar to what was observed experimentally. 

The amounts of slip and the amounts of localization on the most active slip system were used to ascertain how an experiment relates to the ensemble of simulations. For both regions, the experimental data fell within the spread of the values obtained with the DSP model. The adequate agreement demonstrates that at the scale of the performed experiments, the inherent discrete and stochastic nature of plastic deformation needs to be taken into account to capture the correct deformation mechanisms.


\section*{Declaration of Competing Interest}

The authors declare that they have no known competing financial interests or personal relationships that could have appeared to influence the work reported in this paper.

\section*{Acknowledgments}

This research was carried out under project number S17012 in the framework of the Partnership Program of the Materials innovation institute M2i (\href{www.m2i.nl}{www.m2i.nl}) and the Netherlands Organization for Scientific Research (\href{www.nwo.nl}{www.nwo.nl}) (NWO project number 16348).


\bibliography{references.bib}


\end{document}